\documentclass[floats,floatfix,showpacs,amssymb,prd,twocolumn,superscriptaddress,nofootinbib,longbibliography]{revtex4-2}
\usepackage{amssymb,amsmath,verbatim,mathtools,needspace,enumitem,etoolbox,graphicx,physics,microtype,afterpage,xspace,tabularx,lmodern,multirow}
\usepackage{gensymb}
\usepackage{placeins}
\usepackage[dvipsnames, usenames]{xcolor}
\definecolor{linkcolor}{rgb}{0.0,0.3,0.5}
\usepackage[unicode, colorlinks=true, linkcolor=linkcolor, citecolor=linkcolor, filecolor=linkcolor, urlcolor=linkcolor, linktocpage, breaklinks]{hyperref}
\usepackage[all]{hypcap}
\usepackage[T1]{fontenc}
\usepackage[utf8]{inputenc}
\usepackage{mathrsfs}
\usepackage[usenames,dvipsnames]{xcolor}
\usepackage[normalem]{ulem}
\hypersetup{colorlinks=true,citecolor=romared,linkcolor=romared,urlcolor=romared}

\definecolor{romared}{RGB}{142,0,28}

\newcommand{\be}{\begin{equation}}
\newcommand{\ee}{\end{equation}}

\def\be{\begin{equation}}
\def\ee{\end{equation}}
\newcommand{\beq}{\begin{eqnarray}}
\newcommand{\eeq}{\end{eqnarray}}

\usepackage{aas_macros}
\usepackage{makecell}
\usepackage{soul}
\usepackage[nolist,nohyperlinks]{acronym}
\usepackage{orcidlink}
\usepackage{academicons}
\definecolor{orcidlogocol}{HTML}{A6CE39}

\graphicspath{{../Plots/}}

\newcommand{\orcid}[1]{\href{https://orcid.org/#1}{\includegraphics[width=10pt]{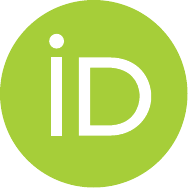}}}

\usepackage{lipsum}

\newcommand{\ben}{\begin{enumerate}}
\newcommand{\een}{\end{enumerate}}

\def\be{\begin{equation}}
\def\ee{\end{equation}}
\def\beq{\begin{eqnarray}}
\def\eeq{\end{eqnarray}}

\def\apjl{\rm{ApJ}}
\def\apjs{\rm{ApJS}}
\def\aap{\rm{A\&A}}

\graphicspath{{./Plots/}}

\allowdisplaybreaks

\begin{document}

\pagenumbering{arabic}

\title{Eternal binaries}
\author{Jaime Redondo-Yuste \orcid{0000--0003--3697--0319}}
\affiliation{Niels Bohr International Academy, Niels Bohr Institute, Blegdamsvej 17, 2100 Copenhagen, Denmark}
\author{Vitor Cardoso \orcid{0000--0003--0553--0433}}
\affiliation{Niels Bohr International Academy, Niels Bohr Institute, Blegdamsvej 17, 2100 Copenhagen, Denmark}
\affiliation{CENTRA, Departamento de F\'{\i}sica, Instituto Superior T\'ecnico -- IST, Universidade de Lisboa -- UL,
Avenida Rovisco Pais 1, 1049--001 Lisboa, Portugal}
\author{Caio F. B. Macedo \orcid{0000--0001--9251--4938}}
\affiliation{Faculdade de f\'{\i}sica, Campus Salin\'opolis, Universidade Federal do Par\'a, 68721-000, Salin\'opolis, Par\'a, Brazil}
\author{Maarten van de Meent \orcid{0000-0002-0242-2464}}
\affiliation{Niels Bohr International Academy, Niels Bohr Institute, Blegdamsvej 17, 2100 Copenhagen, Denmark}
\affiliation{Max Plank Institute for Gravitational Physics, Am Mühlenberg 1, 14476 Potsdam, Germany}
\pacs{}
\date{\today}

\begin{abstract}
The two-body problem is extensively studied in open systems and asymptotically flat spacetimes. However, there are many systems where radiation is trapped: they range from radiating charges in cavities to
low-energy excitations of massive degrees of freedom, to anti-de Sitter spacetimes. Here, we study the problem of motion of a pointlike particle orbiting a massive compact object inside a cavity. We first show that -- assuming circular motion -- there are initial conditions for which the self-force vanishes and the binary is {\it eternal}. We then consider the evolution of the system under radiation reaction in a toy model which we argue captures the essentials of orbiting particles. We show that eternal circular binaries may exist. We also show that the presence of cavity modes leads to chaos in regimes of strong coupling or when the system is initialized close enough to a resonance. Our results have implications for physics in anti-de Sitter spacetimes and possibly for binaries evolving within dark matter haloes, if it consists on massive fundamental fields.
\end{abstract}

\maketitle
\section{Introduction}\label{sec:Introduction}
The problem of motion is foundational for any theory describing a fundamental interaction.
In General Relativity, even the two-body problem -- the description of the motion of a system consisting solely on two pointlike sources -- is a formidable challenge~\cite{Deruelle:1984hq,Poisson:2011nh,Barack:2009ux}. It is made difficult for different reasons, one of them being that dynamical systems emit gravitational waves, which in asymptotically flat spacetimes leads necessarily to an evolution of the system. There are no astrophysically relevant stationary solutions of the two-body problem. 

However, there are relevant instances of confined systems. A particularly interesting example is anti-de Sitter spacetime~\cite{Hawking:1973uf,Hawking:1982dh,Horowitz:1999jd,Cardoso:2006wa}, which has attracted considerable attention after the realization that gravitational physics on the bulk is dual to a field theory living on the boundary~\cite{Maldacena:1997re}.
One can also consider systems which are mostly governed by emission of massive fields. Consider for example a binary of which the components source such a field.
For small orbital frequencies (as compared to the inverse Compton wavelength of the massive field), the excitations of the field remain confined. This example is of more than academic importance, in light of the dark matter challenge. Some proposals advocate the existence of light fields as a possible dark matter component~\cite{PhysRevLett.38.1440,PhysRevLett.40.279,PhysRevLett.40.223,Preskill:1982cy,Abbott:1982af,Dine:1982ah,Arvanitaki:2009fg,Robles:2012uy}. 
A related example concerns dynamics in extra compactified dimensions~\cite{Cardoso:2004zz}, where Kaluza-Klein reduction provides an effective mass to otherwise massless fields. A similar situation occurs also in plasma physics: waves with frequency lower than the plasma frequency are unable to propagate and remain confined within the plasma~\cite{Cardoso:2020nst,Cannizzaro:2020uap}.

Thus, confined systems are important in a number of setups. 
Properties of a radiator inside a cavity have been studied from a quantum and classical 
perspective~\cite{HAROCHE1985347,Haroche_physics_today,DOWLING1991415} when the radiator position is prescribed. It was found that the radiation can be 
extremely suppressed or enhanced depending on the cavity and radiator (in particular, the relative size between 
the cavity and the radiation wavelength is important, and boundary conditions are paramount). But the self-consistent evolution of charges in cavities has not, to the best of our knowledge, been addressed. How does such a system evolve, if it does, under radiation reaction or ``self-force''? 
Here, we wish to take some first steps in this program. We will focus exclusively on a confined system evolving due to the coupling to a scalar field. We will start with a binary, composed of a non-spinning massive compact object at the center, a small orbiting pointlike scalar charge and a confining boundary where Dirichlet conditions are imposed. In regimes where the compact object is very massive (but not a black hole) we can effectively study this situation by considering a massive compact object at the center, with reflective boundary conditions at the surface of the compact object, located at a fixed distance from the Schwarzschild radius. We evaluate the late-time stationary field configurations and compute the (scalar) self-force on the particle, and we also comment on the influence of the initial conditions on the asymptotic state. To actually evolve the particle under radiation reaction, we find the full problem to be still too complex, and we substitute it with a simpler one, which we argue can capture the essentials: a one-dimensional cavity coupled to a harmonic oscillator. We show that in general the state of the oscillator drifts rapidly towards an asymptotic state, and that this drift can be suppressed by appropriately fixing the initial conditions. Moreover, we show that there are regions in parameter space where chaos ensues.  
In what follows we set $G = c = 1$ and greek letters are used to denote spacetime indices $\mu = 0,\dots,3$.

\section{A binary in a cavity}\label{sec:Binary-Cavity}
\subsection{The setup \label{sec:setup}}
%
\begin{figure}
\includegraphics[width = 0.7\columnwidth]{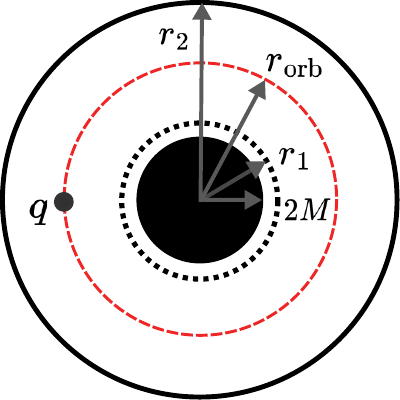}
\caption{Schematic representation of the set-up that we consider: a charge $q$ orbits at a circular orbit of radius $r_{\rm orb}$ a central massive and compact object. There are two mirrors, generating a cavity, located at radius $r_1$ (corresponding to the surface of the star or compact object) and $r_2$.}
\label{fig:Setup}
\end{figure}
We will study a binary system composed of a large massive and compact object, around which a small pointlike object of mass $m_0$ is orbiting on a trajectory $z^\mu(\tau)$. The pointlike object carries a scalar charge $q$ under a field $\Phi$ and the system is described by the action
\beq
S&=&\int d^4x \sqrt{-g} \left(\frac{R}{2k}-\frac{1}{8\pi}g^{\mu\nu}\Phi_{,\mu}\Phi_{,\nu}\right)\nonumber\\
&&\quad-m_0 \int \left(1-\frac{q}{m_0}\Phi\right)\sqrt{-g_{\mu\nu}\dot{z}^\mu \dot{z}^\nu}d\tau\,,
\eeq
where dots are derivatives with respect to the particle's proper time, $k = 8\pi G$ and the above action corresponds to a stress-energy tensor of the pointlike object
\be
T^{\mu\nu}=m_0\int_{-\infty}^{+\infty} \delta^{(4)}(x-z(\tau))\dot{z}^{\mu}\dot{z}^{\nu}d\tau\,,
\ee
and a scalar charge density $\mu$ given by
\be
\mu = q\int d\tau \delta^{(4)}(x-z(\tau))\,,
\ee
where $\delta^{(n)}$ denotes the $n$--th dimensional Dirac's delta. The equations of motion for such system are given by
\beq\label{eq:EoM-general}
\Box \Phi &=& -4\pi \mu, \nonumber\\
\Tilde{m}(\tau)\frac{du^\mu}{d\tau} &=& q(g^{\mu\nu}+u^\mu u^\nu)\Phi_{,\nu}(z)\,, \nonumber \\
\frac{d\Tilde{m}}{d\tau} &=& -q\Phi_{,\mu}(z)u^\mu\,,
\eeq
where the field is evaluated at the trajectory of the charge $z(\tau)$, $u^\mu = \Dot{z}(\tau)$ is the velocity of the particle, and we have promoted the mass of the particle to a dynamical quantity, $\Tilde{m}(z) = m_0 - q\Phi(z)$. Despite the equations involving divergences due to the delta contribution localized at the particle's trajectory, the evolution is perfectly regular: the field can be decomposed in a singular and a regular part in such a way that the latter is the only one responsible for the evolution of the trajectory of the point particle~\cite{Detweiler:2002mi}.  In the following we assume that: (i) the charge to mass ratio of the particle $q/m_0$ is small, so that the backreaction of the field on the particle's trajectory can be studied perturbatively, (ii) the scalar field scales as the charge to mass ratio, in particular, it does not affect significantly the background metric and (iii) the ratio $m_0/M$ between the mass of the particle and the mass of the central object is small, so that we can assume the background geometry to be the Schwarzschild metric:
\beq
g_{\mu\nu}dx^\mu dx^\nu &=& -Ndt^2 + \frac{dr^2}{N} + r^2 d\Omega_2^2\,,\\
N&=&\left(1-\frac{2M}{r}\right)\,,
\eeq
with $d\Omega_2$ the volume form on the $2$-sphere. We will focus on the case in which the particle moves in a circular orbit at some radius $r_{\rm orb}$, the properties of which are dictated by the gravitational pull of the large central mass and by the scalar field acting on the particle. Because we are interested in confined systems, we assume the presence of perfectly reflective mirrors at radius $r_1 > 2M$ and $r_2 > r_{\rm orb} > r_1$. Figure.~\ref{fig:Setup} provides a schematic representation of this setup.

\subsection{Effective source}\label{sec:EffSource}
The scalar field is divergent at the position of the charge. However the self-force acting on the charge due to such a field is finite and scales as $q/m_0$, which is typically small. This apparent contradiction can be explained by noting that the field in a worldtube surrounding the worldline of the charge can be decomposed into two contributions: a singular and a regular part~\cite{Detweiler:2002mi}. The singular part contains the divergences due to the delta contribution in Eq.~\eqref{eq:EoM-general}, and it does \emph{not} contribute to the self-force. On the other hand, the regular part is finite and accounts for the back-reaction of the field on the particle. From the point of view of the equations of motion, the singular part $\Phi_S$ satisfies the equation with the source term $\mu$, whereas the regular part $\Phi_R$ satisfies a homogeneous equation. There have been several different methods developed to compute this regular part. One of them is the so-called \emph{effective source} approach, also referred to as the \emph{puncture method}~\cite{Vega:2007mc, Barack:2007jh}. Intuitively, the idea consists of taking an approximation of the singular field. Then subtracting that approximation from the field itself will result in a term that is equal to the regular part of the field at the location of the particle itself. For general trajectories in curved spacetimes the structure of this singular part is complicated. However by choosing appropriate local Riemann coordinates~\cite{Heffernan:2012su} it can be expanded as 
\begin{equation}
\Phi_S = \frac{q}{\rho}\left(1+A\rho+\dots\right)\,,
\end{equation}
where $A$ is an arbitrary coefficient and $\rho$ is the affine distance to the position of the charge. The effective source method is then based on approximating this by a puncture field, say, 
\begin{equation}\label{eq:Puncture_expansion}
    \Phi_P = q\left(\frac{1}{\rho} + \frac{A_{ijk}}{\rho^3}\Delta x^i\Delta x^j\Delta x^k\right)\simeq \Phi_S\,,
\end{equation}
where $\Delta x^i$ is the coordinate distance in some suitable coordinates~\cite{Heffernan:2012su} and $A_{ijk}$ are coefficients depending on the details of the trajectory. Then, we define the approximate regular field as $\tilde{\Phi}_R = \Phi-W\Phi_P \simeq \Phi_R$, where $W$ is a window function that satisfies $W(\rho) \to 0$ as $\rho \to \infty$, $W(\rho)\to 1$ and $W'(\rho) = W''(\rho) = 0$ as $\rho \to 0$. The regular field now satisfies an inhomogeneous equation of motion given by:
\begin{equation}\label{eq:Effective_Source_Def}
\Box \tilde{{\Phi}}_R=-4\pi \mu-\Box(W \Phi_P)=\mathbf{S}_{\rm eff}\,,
\end{equation}
where the right hand side is referred to as the effective source. This approximate regular field coincides with the actual regular field $\Phi_R$ at the location of the particle itself, but it will not be smooth, in general. A second order (in $\rho$) puncture field will result in a regular field that is $\mathcal{C}^1$ but not $\mathcal{C}^2$ at the location of the charge. However, since the self-force only depends on local properties of the regular field around the position of the particle, a puncture capturing the first two non-trivial orders, such as Eq.~\eqref{eq:Puncture_expansion} is sufficient.

\subsection{Frequency-domain approach to self-force}\label{sec:FD}
We start by computing the self-force on a scalar charge~$q$ orbiting a Schwarzschild exterior of mass $M$ at a circular orbit with radius $r_{\rm orb}$. Instead of the usual (in-going at the horizon and out-going at the outer region) boundary conditions, we consider perfectly reflecting boundary conditions at two radii $r_1$ and $r_2$, satisfying $2M<r_1<r_{\rm orb}<r_2$. We expand the regular part of the field (we drop the tilde from now on and refer to the regular part of the field just as $\Phi_R$) as 
\begin{equation}\label{eq:Field_Expansion_FD}
    \Phi_R = \sum_{\ell=0}^\infty \sum_{m=-\ell}^\ell \int \frac{d\omega}{2\pi}e^{i\omega t} \phi_{\ell m}(r) Y_{\ell m}(\theta, \varphi),
\end{equation}
where $Y_{\ell m}$ are the usual spherical harmonics. To simplify the notation, we will drop the $(\ell m)$ subindex of the field modes $\phi_{\ell m}=\phi$. The equation of motion for each mode on the Schwarzschild background is given by 
\begin{equation}\label{eq:KG_Schwarzschild}
    \begin{aligned}
        \phi_{,rr} + \left(\frac{2}{r} + \frac{N_{,r}}{N}\right)\phi_{,r} + \frac{1}{N^2}(\omega^2-V)\phi = S_{\rm eff}, \\
        V = N\frac{\ell(\ell+1)}{r^2}\,,
    \end{aligned}
\end{equation}
where a comma denotes the partial derivative with respect to the indicated variable. The puncture field for circular orbits in Schwarzschild has been computed to second order in $\rho$~\cite{Warburton:2013lea}:
\beq
\phi_P &=& -\frac{4\pi q}{r_{\rm orb}}Y_{\ell m}(\pi/2, 0) \, \delta(\omega-m\Omega) \,  \frac{g(r)}{r}, \\
g(r) &=&\frac{1}{(2\ell+1)\pi}\sqrt[]{\frac{r_{\rm orb}-3M}{r_{\rm orb}-2M}}\left(2K+\frac{E-2K}{r_{\rm orb}}(r-r_{\rm orb})\right)\nonumber \\
&+&\frac{\lvert r-r_{\rm orb}\rvert}{2r_{\rm orb}(r_{\rm orb}-2M)}\sqrt[]{1-\frac{3M}{r_{\rm orb}}},\label{eq:Puncture_Field}
\eeq
where $E$ and $K$ are complete elliptic integrals of the first and second kind with argument $M/(r_{\rm orb}-2M)$ and  $\Omega^2=M/r_{\rm orb}^3$ is the orbital frequency. We will choose a smooth window function with compact support, similar to the one considered in~\cite{Vega:2007mc}, given by 
\begin{equation}
    \begin{aligned}
        W =& \theta(r-r_{\rm orb} - \sigma)\theta(r_{\rm orb} + \sigma-r) \\
        &\times \mathrm{exp}\left[1- \left(1 - \left(\frac{r - r_{\rm orb}}{\sigma}\right)^4\right)^{-1}\right] ,
    \end{aligned}
\end{equation}
where $\theta(r)$ is the Heaviside step function. The regular field can be obtained then by the method of variation of parameters. Let $\phi_+$ and $\phi_-$ be the solutions to the homogeneous Eq.~\eqref{eq:KG_Schwarzschild} satisfying $\phi_{+}(r_2) =0$ and $\phi_{-}(r_1) =0$. The particular solution is then
\begin{equation}\label{eq:Particular_solution}
    \phi = c_+ \phi_+ + c_- \phi_-,
\end{equation} 
where the coefficients are 
\be
c_+ = \int_{r_1}^r dr \frac{S_{\rm eff}\phi_-}{\text{Wr}(\phi_+, \phi_-)}\, \, \, , \, \, \, c_- = \int_r^{r_2} dr \frac{S_{\rm eff}\phi_+}{\text{Wr}(\phi_+, \phi_-)}\,,
\ee
where $S_{\rm eff}$ is the effective source constructed in Eq.~\eqref{eq:Effective_Source_Def} and $\text{Wr}(\phi_+, \phi_-)= \phi_+\phi_{-}' - \phi_+'\phi_-$ is the Wronskian of the homogeneous solutions. Since the puncture field contains a $\delta(\omega-m\Omega)$ term, only the frequency $\omega=m\Omega$ will contribute to the Fourier expansion of the field. 

However, when the Wronskian of $\phi^{+}$ and $\phi^{-}$ vanishes, the coefficients $c_\pm$ diverge. This happens exactly when $m\Omega$ is one of the \emph{normal mode} frequencies of the system, implying that $\phi^{+}$ and $\phi^{-}$ are the same homogeneous solution. In the case of open boundary conditions this never happens, because all (quasi) normal mode frequencies have strictly positive imaginary part. However, since we have a closed system with reflective boundary conditions, we have an infinite family of normal mode frequencies for each angular mode $\ell$, which we label $\omega_{\ell\, n}$.

The divergence of the coefficients $c_\pm$ is a resonance effect: the orbit is exciting a normal mode of the cavity. In Ref.~\cite{Annulli:2020lyc} it was shown for a simplified situation that the energy in the cavity in the resonant regime increases quadratically with time. In the frequency domain calculation it is assumed that the particle has been at the same orbit for a very long (infinite) period of time. Therefore, it is sensible to expect a divergence in the resonant regime. We will analyze the two different regimes separately. 

Finally, each mode of the self-force $f_a^{\ell m} \equiv f_a$ can be computed from the modes of the regular field like~\cite{Detweiler:2002mi} 
\begin{equation}\label{eq:self-force}
    f_t = \Omega f_\varphi, \quad f_r = \phi_{,r}(r_{\rm orb}) , \quad f_{\varphi} = i m \phi(r_{\rm orb}).
\end{equation}
The self-force $f_{\rm \theta}$ can we always be set to zero by a coordinate redefinition. 
\subsubsection{Non-resonant regime}
\begin{figure}
\includegraphics[width = \columnwidth]{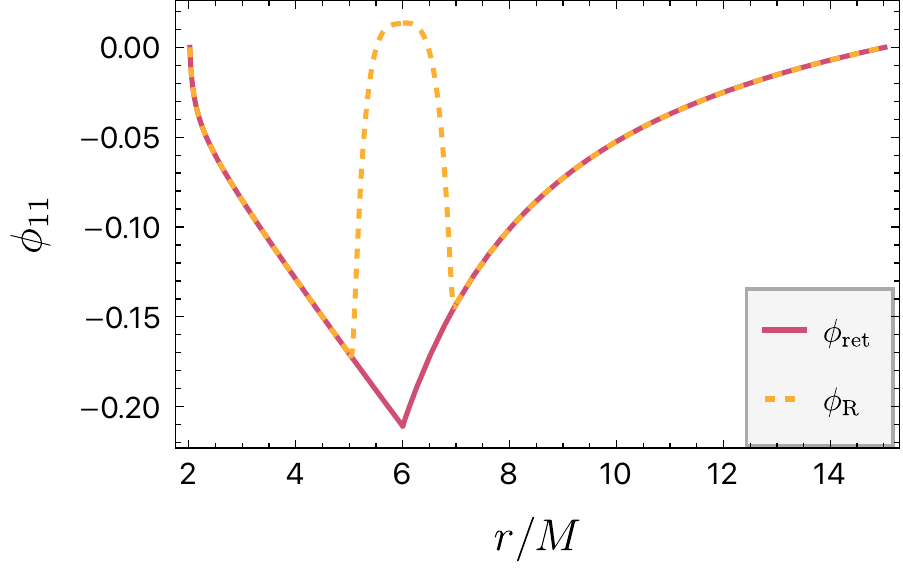}
\caption{Field in the $(1,1)$ mode for a scalar charge orbiting at $r_{\rm orb}=6M$, with mirrors placed at $r_1=2.02M$ and $r_2 = 15M$. The red line is the retarded field, computed from directly integrating the field equation~\eqref{eq:EoM-general}, and the yellow line is the regularized field. Outside of the window function both field coincide.
}
\label{fig:FieldConfig}
\end{figure}

\begin{figure}
   \includegraphics[width = \columnwidth]{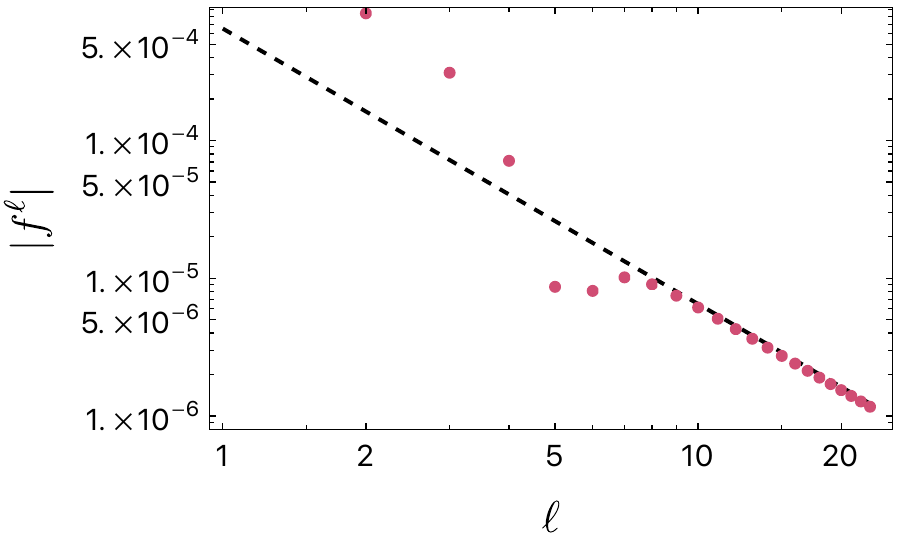}
   \caption{
   	Radial self-force for different $\ell$ modes, for a scalar charge orbiting at $r_{\rm orb}=6M$, with mirrors placed at $r_1=2.02M$ and $r_2 = 15M$. The dashed line represents the $1/\ell^2$ scaling which ensures the convergence of the sum. }
   \label{fig:SFConvergence}
\end{figure}

If $m \Omega \neq \omega_{\ell\, n}$ for all values of $-\ell\leq m\leq \ell$ and all of the normal mode frequencies $\omega_{\ell\, n}$, we can directly integrate~\eqref{eq:KG_Schwarzschild} numerically. The behaviour of the regularized and retarded fields is shown in Fig.~\ref{fig:FieldConfig}. 
From the symmetries of spherical harmonics, the field modes must satisfy the following parity relation:
\begin{equation}
    \phi_{\ell m} = (-1)^m \bar{\phi}_{\ell -m},
\end{equation}
where the bar denotes complex conjugation. The field equation~\eqref{eq:KG_Schwarzschild} is real, as well as the reflective boundary conditions: therefore each field component $\phi_{\ell m}$ is also real. This now guarantees that the above transformation law is just $\phi_{\ell m} = (-1)^m \phi_{\ell -m}$, which in turn is sufficient to show that after summing all the $m$ modes the time and azimuthal angle components of the self-force vanish $F_t = F_\varphi=0$. This is to be expected: the temporal and angular self-force components are related to loss of energy and angular momentum of the charge, respectively. In a closed system, these quantities are conserved and therefore the self-force vanishes. The only non-trivial component is the radial self-force. In Fig.~\ref{fig:SFConvergence} we show that it decays asymptotically like $1/\ell^2$. This guarantees that the resulting self-force converges when summing all of the modes. 

For circular orbits, the radial component of the self-force corresponds to a shift in the orbital frequency of the orbit~\cite{Diaz-Rivera:2004nim}. While the (long-term average of the) orbital frequency is a gauge invariant quantity, the parameterization of the orbit in terms of the radius $r$ is not-gauge invariant~\cite{Detweiler:2008ft}. In principle, we could choose a gauge with a modified radial coordinate $\hat{r}$ such that $\Omega = \hat{r}_{\rm orb}^{-3/2}$ at all orders in perturbation theory. However, this does not mean that the conservative self-force is meaningless. We will discuss a gauge invariant consequence later in Sec.~\ref{sec:ISCOshift}, the ISCO shift.

\subsubsection{Resonant regime}
\begin{figure}
    \centering
    \includegraphics[width = \columnwidth]{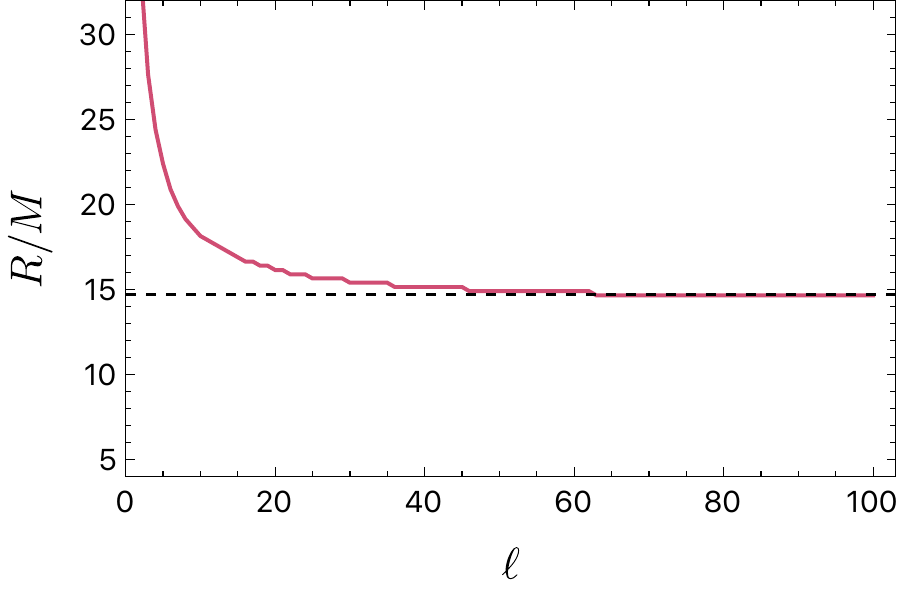}
    \caption{Values at the parameter space at which an orbit at the ISCO radius is first resonant. As $\ell$ increases, the size of the cavity asymptotes to a constant value, represented with a black dashed line. }
    \label{fig:Where_Resonance}
\end{figure}

As we noted, resonances appear when $m \Omega = \omega_{\ell\,n}$. Since the normal mode frequencies $\omega_{\ell\,n}$ form a discrete set, a fine tuning of the orbital radius is required in order to trigger a resonance. However it is interesting to understand what happens physically near this resonant regime. Since the overtone frequencies $\omega_{\ell\,n}>\omega_{\ell\,0}$, any given orbit will first be resonant with the fundamental mode. We have computed the fundamental mode frequencies by solving Eq.~\eqref{eq:KG_Schwarzschild} with a shooting method and observed that at large values of $\ell$, these are well described by a power-law scaling like $\omega_{\ell\,0} \sim \ell / R$, where $R = r_2-r_1$ is the size of the cavity (a coordinate size, but the estimate holds good when the inner boundary is not too close to the Schwarzschild radius).
Therefore large cavities are more likely to trigger resonances, by decreasing the minimum resonant frequency. On the other hand, the maximum orbital frequency is achieved at the innermost stable circular orbit (ISCO), so the critical cavity size $R_c$ such that any cavity with $R \geq R_c$ allows for resonances will scale as
%
$\ell \, \Omega_{\rm ISCO} \sim \frac{\ell}{R_c}$, or $R_c \sim 6^{3/2}M\sim 14.7 \, M$.
%
The precise value of the cavity size for which the ISCO is resonant is shown in Fig.~\ref{fig:Where_Resonance}, for different angular modes $\ell$. As $\ell$ increases the size of the cavity asymptotes to the value of $R_c \sim 15M$, in excellent agreement with the estimate. 

\begin{figure}
    \centering
    \includegraphics[width = \columnwidth]{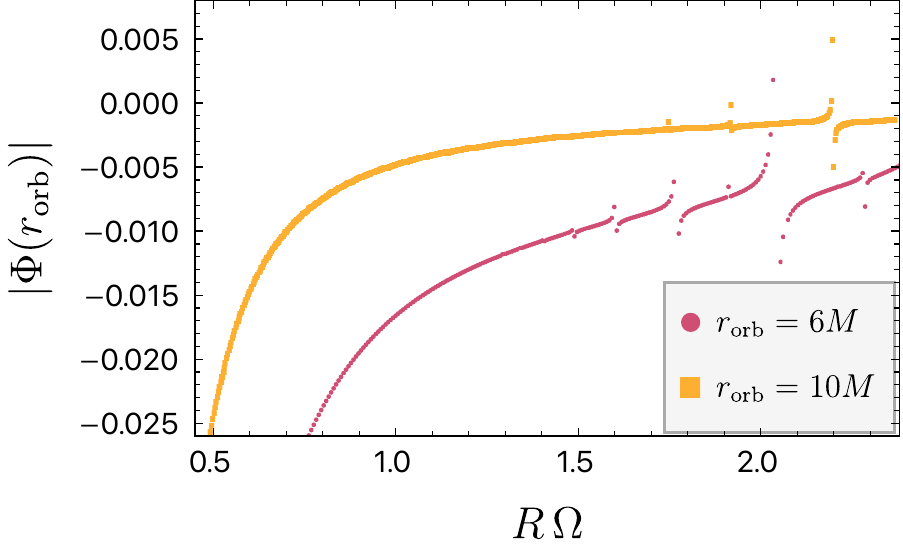}
    \caption{Value of the regularized field at the orbital radius $\Phi_R(r_{\rm orb})$ for $r_{\rm orb}=6M$ (red) and $r_{\rm orb}=10M$ (yellow), as a function of the dimensionless combination $R \Omega$, where $\Omega$ is the orbital frequency. We see that when $R \Omega > 1$, divergences due to orbits become resonant for different angular modes start appearing for both values of $r_{\rm orb}$. }
    \label{fig:Resonance_Behaviour}
\end{figure}

We can understand the behaviour as we approach a resonance by slowly increasing the size of the cavity for a fixed orbital radius. The field at the orbital radius for two different orbits in Fig.~\ref{fig:Resonance_Behaviour}. We see a number of divergences as some modes included in the summation become divergent at different values of the cavity size. We want to emphasize that this divergence is not physical, but is a smoking gun that backreaction is important. 
%
\subsubsection{The ISCO shift}\label{sec:ISCOshift}
A well-known gauge invariant consequence of the conservative piece of the self-force is a shift in the frequency of the ISCO. Recall that the ISCO is defined as the circular orbit located at a vanishing point of the restoring radial force upon perturbations onto slightly eccentric orbits. Under self-force corrections the ISCO frequency will be modified, and can be expanded as
\begin{equation}
	M \Omega_{\rm ISCO} = M\Omega^{(0)}_{\rm ISCO}\left(1 + \frac{q}{m_0} \mathcal{C}_\Omega + \mathscr{O}\left(\frac{q}{m_0}\right)^2\right),
\end{equation}
where $M \Omega^{{(0)}}_{\rm ISCO} = (6M)^{-2/3}$ is the un-corrected ISCO frequency, and $\mathcal{C}_\Omega$ the first-order correction, known as the ISCO shift. This quantity was originally calculated both for the scalar~\cite{Diaz-Rivera:2004nim} and for the gravitational case using self-force results for eccentric orbits~\cite{Barack:2009ey} and this result was also reproduced from the first law of binary mechanics~\cite{LeTiec:2011dp}. The authors of~\cite{Isoyama:2014mja} introduced a way of computing the (gravitational) ISCO shift without studying eccentric orbits, simply starting from a Hamiltonian that reproduces the equations of motion of the point particle. We here generalize their logic to the scalar case. We notice that the equations of motion of the particle~\eqref{eq:EoM-general} coincide with the orbits of the Hamiltonian
\begin{equation}\label{eq:hamiltonian-orbits}
	H = \frac{1}{2m_0}g^{\mu\nu}p_\mu p_\nu\left(1 + \frac{q}{m_0}\Phi(z^\mu)\right) - \frac{q}{2} \Phi(z^\mu),
\end{equation}
where $g^{\mu\nu}$ is the (inverse) Schwarzschild metric, $p_\mu = m u_\mu$ is the particle's four--momentum and $\Phi(z^\mu)$ denotes the field evaluated at the particle's trajectory. The first term accounts for the ``dressed'' mass of the scalar charge $\Tilde{m} = m_0 - q \Phi$, and the second term completes the equations of motion. Note that on-shell, to linear order in $(q/m_0)^2$, the Hamiltonian can be written as a free piece and an interaction term, as 
\begin{equation}
	\begin{aligned}
		H =& H_0 + \frac{q}{m_0} H_1 + \mathscr{O}\left(\frac{q}{m_0}\right)^2, \\
		H_0 =& \frac{1}{2m_0}g^{\mu\nu}p_\mu p_\nu, \\
		H_1 =& \frac{1}{2m_0}g^{\mu\nu}p_\mu p_\nu \Phi(z^\mu) - \frac{m_0}{2} \Phi(z^\mu) \equiv - m_0 \Phi(z^\mu),
	\end{aligned}
\end{equation} 
where the last equality is evaluated on-shell and only valid to linear order in the perturbative parameter $q/m_0$. In this situation, one can follow the logic discussed in~\cite{Isoyama:2014mja} to obtain the ISCO shift in terms of the interaction hamiltonian $H_1$ and the unperturbed redshift function $z_0 = \mathcal{E}_0 - \Omega_0 \mathcal{L}_0$, where $\mathcal{E}_0$ and $\mathcal{L}_0$ are the unperturbed energy and angular momentum of the orbit, respectively. Therefore it is not hard to check that the ISCO frequency shift is given by 
\begin{equation}\label{eq:ISCO-drift}
	\mathcal{C}_\Omega = \frac{z_0 H_1'' + 2 z_0' H_1'}{2 z_0'''},
\end{equation}
where a prime in the above formula denotes derivation with respect to the orbital frequency $\Omega$, and everything is evaluated at $\Omega^{(0)}_{\rm ISCO}$. Since the interaction hamiltonian only depends on the value of the regularized field at the worldline of the particle, the calculation is straightforward. We evaluate $\Phi_R(r_{\rm ISCO} \pm n \Delta r)$, where $\Delta r = 0.01M$ and $n = 0,\dots,5$ and approximate the derivatives by fitting the above data to a polynomial with sufficient order. We have checked that these settings give stable results and that they reproduce the known value of the ISCO shift $\mathcal{C}_\Omega = 0.029$ (c.f.~\cite{Diaz-Rivera:2004nim}) when we consider open out-going boundary conditions. We estimate the error by considering the fit with two polynomials of different order: this uncertainty is typically small, but becomes larger when approaching resonant orbits.
\begin{figure}
	\centering
	\includegraphics[width = \columnwidth]{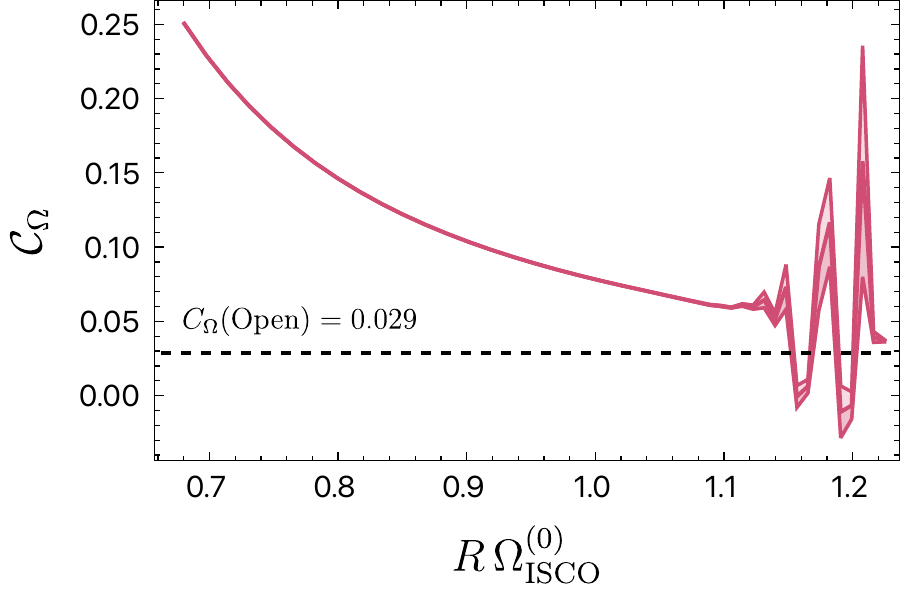}
	\caption{Values of the ISCO shift $\mathcal{C}_\Omega$ as a function of the dimesnionless combination $R \Omega^{(0)}_{\rm ISCO}$. The shaded region takes into account the uncertainty in the calculation, arising from computing the derivatives of the field with respect to the orbital radius numerically. These uncertainties become larger as some of the orbits used in the calculation become resonant with cavity modes. The dashed line represents the value of the ISCO shift when considering open boundary conditions~\cite{Diaz-Rivera:2004nim}. }
	\label{fig:ISCO-shift}
\end{figure}
Figure~\ref{fig:ISCO-shift} shows the ISCO shift for different values of the cavity size. We observe that the shift decreases as the cavity size grows. However, when the cavity is large enough, either the unperturbed ISCO frequency becomes either resonant, or close enough to a resonance, such that the behaviour of the shift becomes oscillatory, and larger uncertainties arise from the numerical calculation of the derivatives in the numerator of Eq.~\eqref{eq:ISCO-drift}. Note that even when $R\to\infty$ the problem is fundamentally different from that of an open system (with outgoing conditions at large distances).

\subsection{Laplace transform approach: including initial conditions}\label{sec:LD}
The field obtained by solving the equation in Fourier space is not a unique solution: one is free to add a solution to the homogeneous equation that satisfies both boundary conditions. The solutions to the homogeneous equation, \emph{i.e.} the normal modes of the cavity, do not decay over time since the system is conservative. These modes can be fixed by specifying initial conditions for the field configuration. Therefore they will have a non-negligible impact in the evolution. 
This is a crucial difference with respect to the open system, since in that case the solutions of the homogeneous equations would be decaying modes. In order to better understand the effect of the initial conditions on the evolution of the binary, we will solve the system using a Laplace transform, which is able to take these into account by introducing minimal modifications to the above calculation. We refer the reader to Appendix~\ref{App:String} for an illustrative comparison between the solution in Laplace and Fourier domain for the case of a $1$-dimensional cavity (a vibrating string). 

The Laplace transform of a field $\phi(t)$ is defined as 
\begin{equation}
\varphi(s) = \int_0^\infty dt \phi(t)e^{-st}.
\end{equation}
Taking the Laplace transform of the Klein-Gordon equation in the Schwarzschild background~\eqref{eq:KG_Schwarzschild} yields the equation
\begin{equation}
\varphi_{,rr}+\left(\frac{2}{r} + \frac{N_{,r}}{N}\right)\varphi_{,r}-\frac{1}{N^2}(s^2+V(r))\varphi = F(s;r),
\end{equation}
where 
\begin{equation}
    F(s;r) = \frac{S(r)}{s-i\Omega}-s\phi_0(r)-\pi_0(r),
\end{equation}
where we denote the initial configuration and momentum of the field by $\phi_0$ and $\pi_0$ respectively. We construct two solutions to the homogeneous equation by imposing the boundary conditions at $r=r_1$ and $r=r_2$, and label them by $\varphi_1(s;r)$ and $\varphi_2(s;r)$, respectively. Now we can write the solution of the problem as
\begin{equation}
    \varphi(s;r) = \int_{r_1}^{r_2}dy F(s;y)G(s;r,y) + \int_{r_1}^r dy F(s;y)H(s;r,y),
\end{equation}
where 
\begin{equation}
    \begin{aligned}
        G(s;r,y) &= \frac{\varphi_1(s;r_2)\varphi_2(s;y)}{ {\rm Wr}(s;r_2)} \frac{\varphi_1(s;r)}{\varphi_1(s;r_2)}, \\
        H(s;r,y) &= \frac{\varphi_1(s;y)\varphi_2(s;r)-\varphi_1(s;r)\varphi_2(s;y)}{{\rm Wr}(s;r)}.
    \end{aligned}
\end{equation}
It is easy to check that this satisfies the equation of motion as well as the boundary conditions. The function $H$ is holomorphic: when $s=i\omega_k$ is one of the normal mode frequencies, the Wronskian in the denominator vanishes, but so does the numerator. However, the function $G$ is meromorphic: there is an additional zero in the denominator due to $\varphi_1(r_2)$. Therefore the situation is analogous to that of a string which we discuss in Appendix~\ref{App:String}. 

We can invert the Laplace transform in the same fashion:
\begin{equation}
    \phi(t,r) = \int_{r_1}^{r_2} dy \int_0^t d\tau f(\tau;y)g(t-\tau;r,y),
\end{equation}
where in this case we have that 
\begin{equation}
    f(t,r) = S_{\rm eff}(r)e^{i\Omega t} - \delta(t)\pi_0(r) - \delta_{,t}(t)\phi_0(r),
\end{equation}
where $\delta_{,t}$ denotes the time derivative of the Dirac's delta. The contribution from the Laplace transform is obtained by summing over the simple poles:
\be
g(t;r,y)= \sum_{k}e^{i\omega_k t}g_k\,,\quad g_k = \frac{\varphi_k(r)\varphi_k(y)}{\rm Wr_{,s}(s=i\omega_k;r_2)}\,,
\ee
where we denote by $\varphi_k$ the eigenfunctions of the homogeneous equation. Now, for a given function $u(r)$, we define its normal mode components as
\begin{equation}\label{eq:nm-modes}
    u_k = \frac{1}{{\rm Wr}_{,s}(s=i\omega_k;r_2)}\int_{r_1}^{r_2} dy u(y)\varphi_k(y),
\end{equation}
Then, the solution is written in a simple form as 
\beq
\phi &=& \sum_k (A_ke^{i\omega_k t} +B_ke^{i\omega_k t} +C_k \left[e^{i\Omega t}-e^{i\omega_k t}\right])\varphi_k(r)\,,\nonumber \\
 A_k &=& \phi_{0 \,k}, \quad B_k = \frac{\pi_{0 \,k}}{i\omega_k}, \quad C_k = \frac{iS_k}{\omega_k-\Omega},
\eeq
where $\phi_{0 \,k}$, $\pi_{0 \, k}$ and $S_k$ are the normal mode components of $\phi_0(r)$, $\pi_0(r)$ and $S(r)$ as defined in Eq.~\eqref{eq:nm-modes}, respectively.
Now the self-force in the $t$ and $\varphi$ directions would vanish (on average during an orbital period) if $\phi(t,r_{\rm orb}) = 0$. This can be easily achieved (in fact, for all $r$) just by requiring, e.g. the following initial configuration of the field:
\begin{equation}\label{eq:FT-Condition-Laplace}
 \phi_{0 \, k} = \frac{i S_k}{\omega_k-\Omega}, \quad \pi_{0 \, k} = 0.
\end{equation} 
This initial field configuration is just a configuration adapted to the regularized field at the position of the charge. This way, the energy (angular momentum) that the charge absorbs from this initial configuration exactly compensates the energy (angular momentum) emitted. Notice how this condition is very similar to the one that would be obtained by applying the same procedure to a $1$-dimensional cavity (a vibrating string), as is discussed in Appendix~\ref{App:String}.

Writing this solution, we can also analyze what happens in the resonant regime. Assume, without loss of generality, that the system is resonant at the fundamental mode $\omega = \omega_0$. Taking the limit $\Omega \to \omega_0$ yields a regular solution
\begin{equation}
    \begin{aligned}
        \phi = &\sum _{k\neq 0} (A_ke^{i\omega_k t} +B_ke^{i\omega_k t} +C_k \left[e^{i\Omega t}-e^{i\omega_k t}\right])\varphi_k(r) \\
        & + S_0 t e^{i\omega_0 t} \varphi_0(r).
    \end{aligned}
\end{equation}
The amplitude grows linearly in time due to the resonant modes in the source term $S_0$ (c.f. the case of the vibrating string in Appendix~\ref{App:String}). This corresponds to a quadratic growth in the energy, which is consistent with~\cite[]{Annulli:2020lyc}. For the resonant case it is not possible to cancel the self-force: even if we could cancel it over a period, eventually it would be large enough that averaging over a period would be meaningless. Since the self-force will also grow linearly in time, it would be necessary to take into account the evolution of the orbit.

\section{Including backreaction: a toy model}

In the previous analysis, it is implicitly assumed that the self-force acting on the charge does not modify its trajectory. However, in a realistic case, the dynamics of the field and the particle are coupled in a non-trivial way. Consistently evolving the field and the trajectory of a radiating particle is a long-standing problem that poses deep challenges. Briefly speaking, the main issue is that the coupled systems of equations is ill-posed. Schematically the problem is the following:
\be
\Box\Phi = S_{\rm eff}(x, \dot{x}, \ddot{x}, \dddot{x}, \dots)\,,\qquad \ddot{x} = \nabla\Phi.
\ee
The source term depends upon derivatives of the acceleration. Since there are only second order equations for the trajectory of the particle, the system, as written, is ill-posed. This is the same issue already present in the Abraham-Lorentz equation. Both perturbative and reduction of order schemes have been proposed to address this problem~\cite{Gralla:2009md, Lanir:2017hwl}. The currently standard approach to evolve such systems is to employ a two-timescale expansion~\cite{Hinderer:2008dm,Pound:2021qin} separating orbital and evolutionary timescales. Attempts at directly solving the self-consistent equations in the time-domain have only been partially successful~\cite{Diener:2011cc, Heffernan:2017cad}, and remain an open problem. 

Here we do not attempt either approach to solve the problem. However we expect the self-force to be sufficiently small so that approximations can give reasonably good results for the setting of our interest. In particular, we can consider that the charge moves from one circular orbit to another, having effectively a single degree of freedom: the orbital frequency $\Omega$. The value of $\Omega$ would then evolve slowly depending on the self-force. More crucially, a particular characteristic of the system is that, since it is enclosed on a cavity, it conserves the total energy (and angular momentum, as well). 
Therefore we will construct a toy model which is a Hamiltonian system, such that the energy conservation is guaranteed, and that captures some of the physical characteristics of the self-force problem. We will use this toy model to explore whether there are really eternal binaries, \emph{i.e.} configurations such that the orbital frequency remains bounded within a certain range (\emph{e.g.} it is natural to require the frequency to be $\Omega \leq \Omega_{\rm ISCO}$). Moreover we will also use this system to explore what happens when the system is initialized close to a resonant orbit, and discuss the regimes in which the orbital motion will become chaotic. 

\subsection{Hamiltonian model}
We consider a $1$-dimensional cavity of size $L$, with canonical variables $(\phi,\,\pi)$ coupled to a harmonic oscillator (with action-angle variables $q$, $p$). Their dynamics is governed by the hamiltonian:
\begin{equation}
    \begin{aligned}
        H =& \frac{p^2}{2} + \frac{1}{2L}\int_0^L dx (\pi^2 + \phi_{,x}^2)\\
        &-\frac{\epsilon}{L} \cos(q/L) \int_0^L dx \frac{\phi(x)}{L}S(x),
    \end{aligned}
\end{equation}
where $\epsilon$ is a coupling parameter and $S(x)$ a coupling function. The Hamilton equations of the system are just 
\begin{equation}\label{eq:toy-model-first-order}
    \begin{aligned}
        \phi_{,t} &= \pi, \quad \pi_{,t} = \phi_{,xx} + \frac{\epsilon}{L} \cos(q/L) S(x), \\
        q_{,t} &= p, \quad p_{,t} = - \frac{\epsilon}{L^2} \sin(q/L) \int_0^L dx \frac{\phi(x)}{L}S(x). 
    \end{aligned}
\end{equation}
Notice that the conjugate momentum $p$ is mapped to the angular frequency of the oscillator, which we sometimes refer to as $\Omega$.
In second order form these equations are just 
\begin{equation}
    \begin{aligned}
        \phi_{,tt} - \phi_{,xx} &= \frac{\epsilon}{L} \cos(q/L) S(x), \\
        q_{,tt} &= - \frac{\epsilon}{L^2} \sin(q/L) \int_0^L dx \frac{\phi(x)}{L}S(x). 
    \end{aligned}
\end{equation}
Therefore, the oscillation of the cavity is sourced by the harmonic oscillator, which then experiences a back-reaction depending on the cavity configuration. Since the Hamiltonian does not explicitly depend on time, the total energy of the system given by $H$ is conserved, as can also readily be confirmed explicitly from the equations of motion.

\subsection{Perturbative calculation}
In this section we derive some analytical results for the above system. To take the system to a simple form, we expand the cavity in its normal modes, 
\begin{equation}
    \phi = \sum_k c_k \phi_k ,
\end{equation}
where the $\{\phi_k\}$ are orthonormal and their associated normal frequencies are $\omega_k$. We also expand the momentum as $\pi = \sum_k b_k \phi_k$. Then, integrating by parts and using the equations of motion, the Hamiltonian becomes
\begin{equation}
    H = \frac{p^2}{2}+\frac{1}{2L}\sum_k (b_k^2 + \omega_k^2 c_k^2)-\frac{\epsilon}{L} \cos(q/L) \sum_k \frac{c_k}{L} s_k,
\end{equation}
where 
\begin{equation}
s_k = \frac{1}{L^2}\int_0^L dx S(x)\phi_k(x).
\end{equation}
For simplicity we assume that $s_k = \delta_{k \, k_0}$ for some $k_0$. Then we have a collection of infinite oscillators with mode number $k \neq k_0$, and two coupled oscillators. The reduced coupled system is just 
\begin{equation}
    H = \frac{L^2}{2}(b^2 + \omega^2 c^2) + \frac{p^2}{2} - \epsilon c \cos(q/L),
\end{equation}
where we drop the $k_0$ sub-index for simplicity. We can take one step further and rewrite the degrees of freedom of the cavity mode in action-angle variables (of the decoupled system):
\begin{equation}\label{eq:action-angle-transformation}
    c = \sqrt[]{\frac{2J}{\omega}}\cos\varphi, \quad b = \sqrt[]{2\omega J}\sin\varphi,
\end{equation}
so the hamiltonian reduces to 
\beq
H &=& H_{\rm free} + \epsilon h, \\
H_{\rm free} &=& L^2 \omega J + \frac{p^2}{2},\, h= - \, \sqrt[]{\frac{2J}{\omega}} \cos\varphi \cos(q/L)\,.
\eeq
In what follows, we will set $L = 1$ for simplicity (we can recover the length units using the starting Hamiltonian as guiding principle whenever necessary). The perturbation is $2\pi$-periodic in the angles $\{\varphi, q\}$. Therefore this system is written in standard form, according to Ref.~\cite{kevorkian2012multiple}, which allows us to find a perturbative solution using near-identity transformations. We will find a near-identity transformation that maps the current canonical degrees of freedom $(p_a, q^a) \mapsto (P_A, Q^A)$ such that the new interaction hamiltonian is trivial $\mathcal{H} = H_{\rm free} + \mathscr{O}(\epsilon)^3$. The generating function $F(q^a, P_a, \tilde{t}) = q^a P_a + \epsilon F_1 + \epsilon^2 F_2 +\mathscr{O}(\epsilon^3)$ (where $\tilde{t} = \epsilon t$) is obtained in Appendix~\ref{App:Transformation}. We consider, for simplicity, that the initial state is 
\begin{equation}
(q_1, q_2, p_1, p_2)\rvert_{t = 0} = (0, q_{2 (0)}, \Omega_0, p_{2 (0)}),
\end{equation}
here $q_{2 (0)}$ and $p_{2 (0)}$ are obtained by applying the inverse of the transformation~\eqref{eq:action-angle-transformation} to the initial conditions for the cavity, $c_{(0)} = y_0$ and $b_{(0)} = 0$. The solution in the new variables $(Q_i, P_i)$ is trivial, 
\begin{equation}\label{eq:Perturbative-Sol-New-Vars}
\begin{aligned}
P_i(t) &= P_i(0) + \mathscr{O}(\epsilon^3),\\
Q_1(t) &= P_1(0) t + Q_1(0)+ \mathscr{O}(\epsilon^2), \\
Q_2(t) &= \omega t + Q_2(0)+ \mathscr{O}(\epsilon^2).
\end{aligned}
\end{equation}
So we only need to use the generating function to obtain the initial conditions in the transformed variables. The transformation is given by
\begin{equation}
    \begin{aligned}
        P_i &= p_i - \epsilon \frac{\partial F_1}{\partial q_i} - \epsilon^2 \left(\frac{\partial F_2}{\partial q_i} - \frac{\partial^2 F_1}{\partial q_i \partial P_j}\frac{\partial F_1}{\partial q_j}\right)+ \mathscr{O}(\epsilon^3), \\
        Q_i &= q_i + \epsilon \frac{\partial F_1}{\partial P_i} + \epsilon^2\left(\frac{\partial F_2}{\partial P_i} - \frac{\partial^2 F_1}{\partial P_i \partial P_j}\frac{\partial F_1}{\partial q_j}\right)+ \mathscr{O}(\epsilon^3).
    \end{aligned}
\end{equation}
Using that transformation, it is straightforward to obtain the initial configuration in the new $(Q_i, P_i)$ variables. The evolution in the original values is obtained by taking the inverse transformation, evaluated at the solution~\eqref{eq:Perturbative-Sol-New-Vars}:
\begin{equation}
    \begin{aligned}
        p_i &= P_i + \epsilon \frac{\partial F_1}{\partial q_i} + \epsilon^2 \left(\frac{\partial F_2}{\partial q_i} - \frac{\partial^2 F_1}{\partial q_i \partial q_j}\frac{\partial F_1}{\partial P_j}\right)+ \mathscr{O}(\epsilon^3), \\
        q_i &= Q_i - \epsilon \frac{\partial F_1}{\partial P_i} - \epsilon^2\left(\frac{\partial F_2}{\partial P_i} - \frac{\partial^2 F_1}{\partial P_i \partial q_j}\frac{\partial F_1}{\partial P_j}\right)+ \mathscr{O}(\epsilon^3).
    \end{aligned}
\end{equation}
We can use the resulting transformed momentum $p_1$ to connect back to the physical quantities of interest. We are in particular interested in the asymptotic value of the momentum $p_1(t \to \infty) = \Omega_\infty$, which is given by
\begin{equation}
    \Omega_\infty = \lim_{t\to\infty}\frac{1}{t}\int_0^t p_1(t')dt'.
\end{equation}
Now computing this integral is in general complicated: the near-identity transformation allows us to simplify the solution in the transformed variables $(Q_i, P_i)$, but not so much in the original variables $(q_i, p_i)$. However, we can use the fact that the only time-dependence of the momentum $p_1$ is through the variables $Q_i$, so that $p_1(t) = p_1(Q_1(t), Q_2(t))$. Now, when taking the limit as $t\to\infty$, since $Q_i(t) = \omega_i t + \dots$, we can expect that outside of resonant regimes the system will explore all of the possible configurations in the torus $(Q_1, Q_2)$, so that the integral is ergodic. A critical observation is that the original hamiltonian is periodic in the ``angle'' variables, since this guarantees that the corresponding momenta (and therefore the frequencies $\omega_i$ that determine the evolution of the $Q_i$) are fixed. Therefore we can replace the limit by the average values over the torus like (see also~\cite{Drasco:2003ky})
\begin{equation}
    \Omega_\infty = \int_0^\infty dt' \sum_{mn} \Tilde{p}_1^{mn} e^{i(m \omega_1 + n \omega_2)t},
\end{equation}
where $\omega_1 = P_1(0)$, $\omega_2 = \omega$ and 
\begin{equation}
    \Tilde{p}_1^{mn} = \frac{1}{4\pi^2}\int dQ_1 dQ_2 p_1(Q_1, Q_2) e^{-i(mQ_1+ nQ_2)}.
\end{equation}
Finally doing the time integral means that only the $\tilde{p}_{00}$ term contributes, so that the final result is 
\begin{equation}
    \Omega_\infty = \Tilde{p}_1^{00} = \frac{1}{4\pi^2}\int dQ_1 dQ_2 p_1(Q_1, Q_2).
\end{equation}
This now can be easily evaluated, obtaining
\begin{equation}\label{eq:Drift_As}
    \begin{aligned}
        \Omega_\infty =& \Omega_0 - \epsilon \frac{\Omega_0 y_0}{\Omega_0^2 - \omega^2}-\epsilon^2 \frac{1}{4\Omega_0} \Big( 6 + 8y_0^2   \\
        & - \frac{\omega^2(8+19y_0^2)}{\omega^2-\Omega_0^2} + \frac{10\omega^2 y_0^2 (\omega^2 + \Omega_0^2)}{(\omega^2 - \Omega_0^2)^2}\Big) 
        + \mathscr{O}(\epsilon^3).
    \end{aligned}
\end{equation}
This expression captures the main results of the dynamical evolution of the system: (i) the frequency of the oscillator \emph{drifts} from its initial value $\Omega_0$ by an amount which is supressed by the coupling constant, (ii) the leading order contribution to the drift is due to the initial content in the cavity ($y_0$), which would drive the exchange of energy between cavity and oscillator. If the cavity is initially un-excited, then the oscillator first needs to populate the cavity modes to which it couples, and then that same coupling would drive the drift, but this ``self-coupling'' only appears at second order, as expected. (iii) The drift diverges as $\Omega_0 \to \pm \omega$, i.e. as we approach a resonant state. Notice that our perturbative analysis is not valid at resonances: the near-identity transformation diverges at the resonances (c.f. Eq~\eqref{eq:F2-NIT}) and therefore cannot be assumed to be close enough to the identity anymore.

It is interesting to highlight a particularly interesting set-up, which is the case where the cavity contains initially an excitation which is of the same order of magnitude as the coupling, $y_0 = \epsilon \phi_0$. In this case, the asymptotic value of the drift $D_\infty= \Omega_\infty - \Omega_0$ is given, up to second order in the perturbative parameter, by
\begin{equation}\label{eq:Drift_As_Expanded}
    D_\infty = \epsilon^2\frac{4\phi_0\Omega_0^2(\omega^2-\Omega_0^2) -\omega^2-3\Omega_0^2}{4\Omega_0(\Omega_0-\omega)^2(\Omega_0+\omega)^2}.
\end{equation}
Therefore it is always possible to find some initial excitation $\phi_0$ such that the asymptotic value of the drift vanishes. This ``fine-tuned'' initial condition is given by 
\begin{equation}\label{eq:fine-tuned-no-drift}
    \phi_{0}^{\rm FT} = \frac{3\Omega_0^2 + \omega^2}{4\Omega_0^2(\omega^2 - \Omega_0^2)}.
\end{equation}
Interestingly this does not coincide with the fine-tuned initial condition predicted in Eq.~\eqref{eq:FT-Condition-Laplace}. However recall that the condition obtained depends very precisely on the details of the coupling term between the cavity and the oscillator (the particle's trajectory, in the full self-force case). In this case, we have considered several simplifications that allow us to compute the value of the initial configuration modes of the cavity just in terms of the parameters of the system. However, both results coincide in requiring that the initial configuration of the cavity is supressed by $\epsilon$ (by the scalar charge to mass ratio in the self-force case), and that it diverges if the system is initialized at a resonance. 

\subsection{Numerical Solution}
The reduced model that we consider to study the backreaction allows, in principle, for more complicated source terms than the one that we have considered for the perturbative scheme. In fact, the typical structure of the effective source for the self-force problem is that of a narrow pulse, which will generally excite a large number of modes of the cavity. Studying the interplay of different modes of the cavity coupling to the same oscillator is challenging, from the analytical point of view. However, we can explore whether that introduces additional physical features by solving the system numerically. We refer the interested reader to Appendix~\ref{App:Numerics} for a description of the numerical methods used. 

In this section we first test the accuracy of our numerical scheme by comparing the solution obtained numerically with the perturbative solution for the case where the coupling function just couples a single mode. Then we consider a more realistic case by studying a sinusoidal coupling, $S(x) = \sin x$. Since the normal modes of the cavity can be written as trigonometric functions, this allows for a very simple implementation of the interaction term while producing the desired coupling between several modes of the cavity. 

\begin{figure}
	\centering
	\includegraphics[width = \columnwidth]{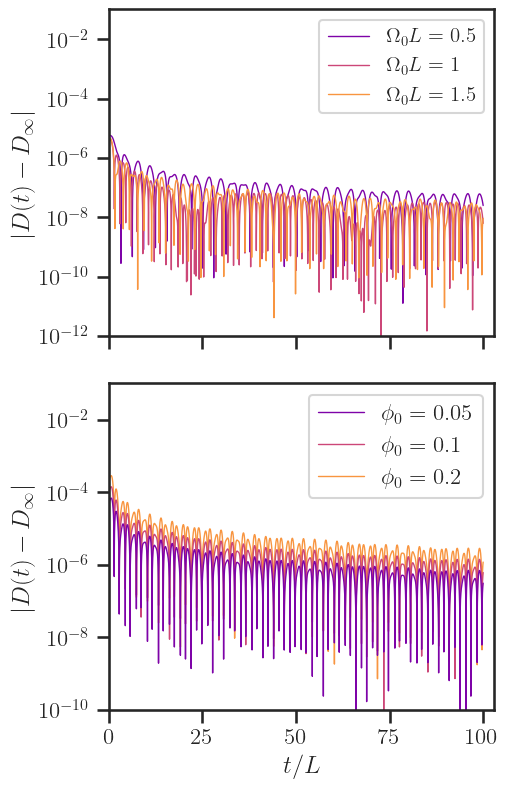}
	\caption{Evolution of the drift $D(t)$ defined in Eq.~\eqref{eq:drift-defn} for different values of $\Omega_0 L$ (top) with $y_0 = 0$ and for different values of $y_0$ (bottom) while keeping $\Omega_0 L = 1$, with respect to the asymptotic value~\eqref{eq:Drift_As}. In both situations we fix the perturbative parameter $\epsilon = 0.01$. }
	\label{fig:ComparisonAnalytics1}
\end{figure}
%

\subsubsection{Comparison with perturbative solution}
First we choose a source term which is just the fundamental mode of the cavity, so that we can compare explicitly our numerical solution with the analytical predictions using perturbation theory. We evolve the system numerically and compute the drift at any given time. We define the local value of the drift as 
\begin{equation}\label{eq:drift-defn}
    D(t) = \frac{1}{t}\int_0^t (\Omega(t')-\Omega_0)dt'.
\end{equation}
We show its behaviour for different values in parameter space in Fig.~\ref{fig:ComparisonAnalytics1}, where we compare it with the asymptotic value predicted by Eq.~\eqref{eq:Drift_As}. The result clearly shows that the drift approaches very quickly the value predicted by perturbation theory, with the difference being of $\mathscr{O}(\epsilon^3)$, consistently with the fact that the perturbative solution $D_\infty$ is only valid to second order. 

\begin{figure}
    \centering
    \includegraphics[width = \columnwidth]{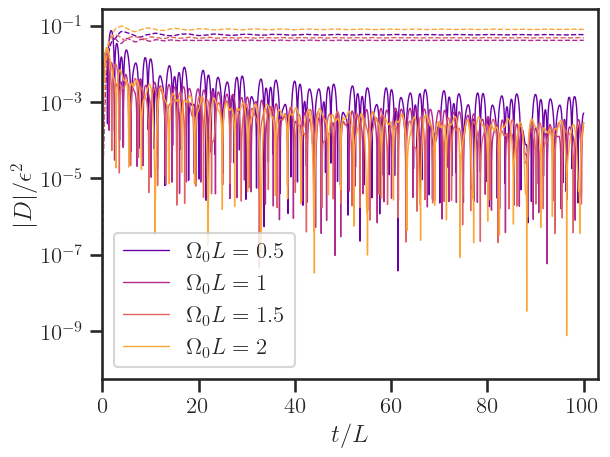}
    \caption{Evolution of the drift $D(t)$ for different values of $\Omega_0 L$ as indicated in the legend, and for the fine-tuned initial conditions defined in Eq.~\eqref{eq:fine-tuned-no-drift}. The dashed lines represent the same situation, but with trivial initial conditions $y_0 = 0$. We observe that the drift, which usually enters at $\mathscr{O}(\epsilon^2)$, becomes a higher order effect due to the initial conditions.}
    \label{fig:ComparisonAnalytics2}
\end{figure}

Secondly, we test whether the prescription to vanish the frequency drift~\eqref{eq:fine-tuned-no-drift} actually suppresses the drift. We represent the evolution of the drift obtained numerically using $y_0 = y_0^{\rm FT}$ as initial condition for different values of $\Omega_0 L$. We observe in Fig.~\ref{fig:ComparisonAnalytics2} that the drift is very supressed: the normalized value of $D(t)/\epsilon^2$ becomes approximately 2 orders of magnitude smaller by fine-tuning the initial conditions, which is consistent with a residual drift of order $\mathscr{O}(\epsilon^3)$, since we are setting $\epsilon = 0.01$ in the figure. It would be sensible to assume that by increasing the perturbative order we could improve the prescription for the fine-tuned initial conditions and suppress the drift even further. 
%

\subsubsection{Multiple mode excitation}
In the case studied perturbatively the system has only $2$ coupled degrees of freedom. However in a more realistic scenario the interaction term will couple multiple modes of the cavity to the oscillator. Even though nothing changes fundamentally for the perturbative analysis, the increased number of degrees of freedom complicates the calculation. However, we can explore whether the coupling to multiple degrees of freedom has any effect in the physics by numerically solving the equations, with coupling function $S(x) = \sin x$. 

\begin{figure}
    \centering
    \includegraphics[width = \columnwidth]{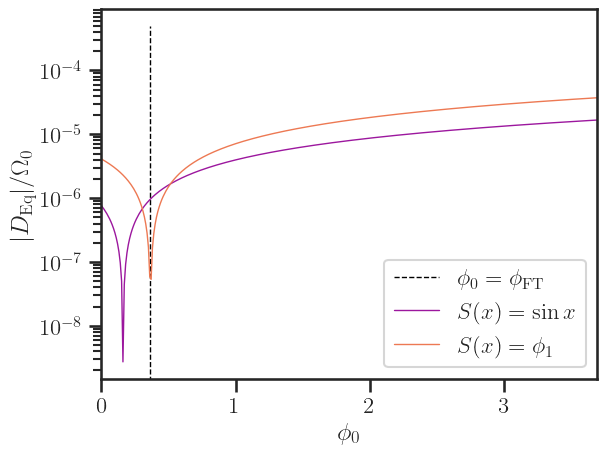}
    \caption{Asymptotic value of the drift $D_{\rm Eq}$ as a function of the initial configuration of the cavity $\phi_0$. The black line denotes the value predicted by Eq.~\eqref{eq:Drift_As}, and the purple and orange lines consider the fundamental mode or $S(x) = \sin x$ as coupling functions, respectively. We set $\epsilon = 0.01$ in this case.  }
    \label{fig:driftVsY}
\end{figure}

First, we observe that as predicted the asymptotic value of the drift can be made arbitrarily small by fine-tuning the initial conditions. Let us choose an initial profile given by the fundamental mode, with amplitude $\phi(t=0) = \epsilon \, \phi_0$. Then, we see the reduction in the asymptotic value of the drift as a pronounced valley in Fig.~\ref{fig:driftVsY}. The location of this dip is exactly where predicted by Eq.~\eqref{eq:Drift_As} when we choose the fundamental mode as coupling function. When we choose a different coupling function, this peak gets displaced slightly, as one would naturally expect. However, Eq.~\eqref{eq:Drift_As} is still a reasonable approximation for the initial conditions necessary to make the asymptotic drift vanish. 

\begin{figure}
    \centering
    \includegraphics[width = \columnwidth]{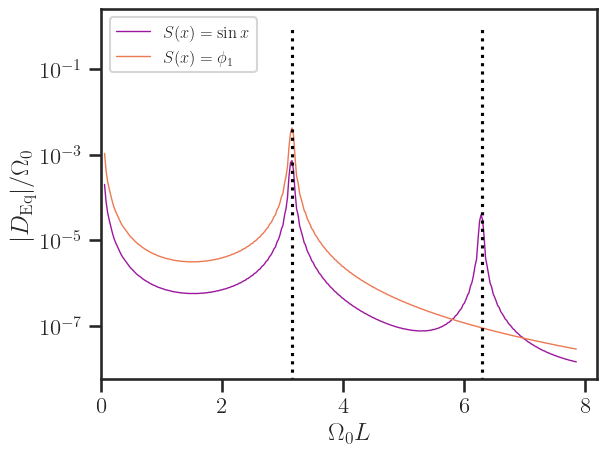}
    \caption{Asymptotic value of the drift $D_{\rm Eq}$ while changing the initial frequency $\Omega_0 L$ for the two choices of coupling functions considered (as indicated in the legend). We represent as dotted black lines the resonant frequencies, \emph{i.e.} the normal modes of the cavity. }
    \label{fig:driftVsW}
\end{figure}

So far it would seem that the difference between the two choices considered here for coupling functions is just quantitative. Although this is a valid observation, it is only correct in a particular range of parameter space. We observe this clearly in Fig.~\ref{fig:driftVsW}. By choosing a coupling function that excites multiple normal modes of the cavity, there are new resonances, for example, at $\Omega_0 L= 2\pi$, which were absent both in the perturbative solution or in the numerical solution using the fundamental mode as a coupling function. However we also note that in the range $\Omega_0 L \in [0, \omega]$ both coupling functions result in a qualitatively similar behaviour. 

\begin{figure}
    \centering
    \includegraphics[width = \columnwidth]{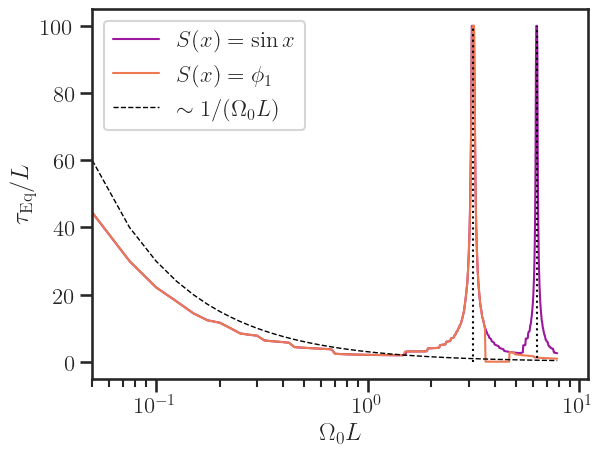}
    \caption{Equilibrium timescale $\tau_{\rm Eq}$ as a function of the initial frequency $\Omega_0 L$ for both choices of coupling functions (see legend). When $\Omega_0 L\ll 1$, the equilibrium time scales as $1/\Omega_0$ consistently with the perturbative analysis.}
    \label{fig:TEQ}
\end{figure}

Finally, we want to explore how quickly the system achieves equilibrium in both cases: for this purpose we define the equilibrium timescale $\tau_{\rm Eq}$ as the time at which the frequency completes the first oscillation. We show our numerical results in Fig.~\ref{fig:TEQ}, which show a very good agreement between both coupling functions in the regime where the coupling to a single mode dominates. We observe that far from resonances this equilibrium timescale scales as $\tau_{\rm Eq}\sim 1/\Omega_0$. This can be understood directly from the method we used to obtain the asymptotic value of the drift in the perturbative analysis: the transformed canonical angles $Q_i$ evolve with frequencies $\omega_i = (P_1, \omega)$. The time-scale it takes for each angle to cover the whole torus is therefore on the scale $\tau \sim 1/\min(\omega_i)$. To leading order, $P_1 = \Omega_0$, so when $\Omega_0 L \ll 1$ this will be the frequency which dominates the time-scale of the system, resulting in the observed scaling. As the system approahces the resonance, the frequency of the fundamental mode becomes the one that dominates the analysis. In particular, exactly at the resonances the trajectories that the system explores in phase space do not cover the whole torus and therefore equilibrium is never achieved: this explains the divergence observed in Fig.~\ref{fig:TEQ} at the resonant frequencies. 

%

\subsubsection{Chaotic orbits}
A natural feature of coupled oscillators is the presence of chaos. This characterizes a regime where small variations in the initial conditions lead to large differences in the dynamical evolution. In such a regime the system will need, in general, take an arbitrarily large time to relax to equilibrium. We will combine observations from the numerical solution with results from perturbation theory to characterize the transition towards chaos in the system considered here.

\begin{figure}
    \centering
    \includegraphics[width = \columnwidth]{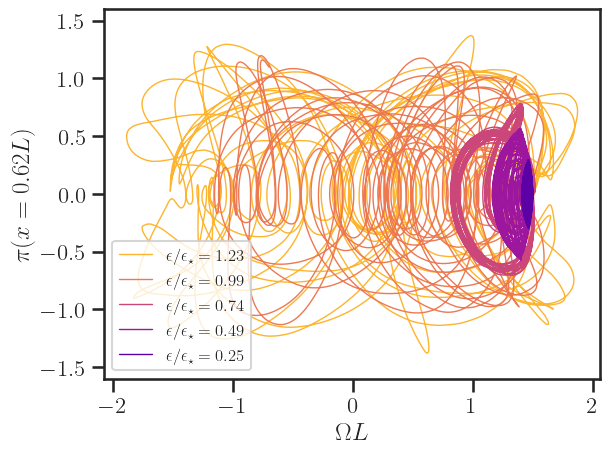}
    \caption{Trajectory in momentum space $(\Omega, p)$ for different values of the perturbation scale $\epsilon$, normalized with the critical perturbative parameter $\epsilon_\star$ defined in Eq.~\eqref{eq:epsilon-crit}. We observe that when $\epsilon > \epsilon_\star$ the system becomes chaotic, exploring a significantly larger fraction of phase space. }
    \label{fig:phasespace}
\end{figure}

In order to explore this, we show in Fig.~\ref{fig:phasespace} the trajectory in a projection of phase space (in the momentum plane). By increasing $\epsilon / L$ the system starts to explore a larger portion of phase space. Eventually, when $\epsilon \gg L$ the system transitions towards a chaotic regime. In this scenario, the global amplitude of the oscillations in the momentum coordinate $\Omega$ is comparable to the distance between resonances of the system, $\Delta \Omega \sim \delta \omega$. This motivates a temptative definition of a critical value for the perturbative scale $\epsilon_\star$ (at a fixed value of $\Omega_0$) such that the system becomes chaotic. If we assume the validity of Eq.~\eqref{eq:Drift_As} beyond perturbation theory, and letting $\phi_0 = 0$ for simplicity, we require: 
\begin{equation}
    \frac{\epsilon^2}{L^4}\frac{\omega_n^2 + 3\Omega_0^2}{4\Omega_0(\Omega_0^2-\omega_n^2)^2} = \omega,
\end{equation}
where for this system $\delta \omega = \omega$ the distance between resonances coincides with the frequency of the fundamental mode. In the above expression, $\omega_n$ is the frequency of the normal mode closest to $\Omega_0$. This implies that 
\begin{equation}\label{eq:epsilon-crit}
    \epsilon_\star/L^2 = \frac{2\sqrt{\Omega_0\omega} \lvert \Omega_0^2 - \omega_n^2\rvert}{\sqrt{\omega_n^2 + 3\Omega_0^2}}.
\end{equation}
Remarkably, this estimate seems to estimate correctly the transition towards chaos observed numerically, even when we consider the excitation of multiple modes by using a sinusoidal coupling $S(x) = \sin x$, as shown in Fig.~\ref{fig:phasespace}.
%

\section{Consequences for binary systems}

The toy model that we have explored in the previous section, while highly simplified, shares many features with the self-force problem that we aimed to explore in the first place. In this section we discuss how the main conclusions extracted from the toy model hold for the more realistic set-up. 

Firstly, the system discussed in section~\ref{sec:setup} is conservative. It is possible to write a Hamiltonian whose orbits coincide with the trajectories of the particle~\eqref{eq:hamiltonian-orbits}, and the scalar field can also be described by a Hamiltonian. Ignoring the spin of the point-particle, its dynamics can be described by a hamiltonian in action-angle variables for two oscillators~\cite{Hinderer:2008dm, Fujita:2016igj}. The main difference, then, is that we replace a single harmonic oscillator with two more complicated oscillators. In a similar manner, the scalar field dynamics is more complicated: on the one hand, each harmonic mode $(\ell,m)$ evolves independently in a slightly different way and is sensitive to a different set of resonant frequencies, and on the other hand, the wave equation introduces the Schwarzschild potential. In the linear regime the different modes do not couple so we can consistently study their evolution independently. Moreover, as discussed in Appendix~\ref{App:Flat} the dynamics of the field does not depend crucially on the potential: its effect reduces to changing the spectrum of normal modes. Our results only depended in the property that any field configuration can be expanded as a sum over these normal modes, which form an orthonormal basis. Since this is still valid, the hamiltonian term used to describe the one-dimensional vibrating cavity captures all the relevant physics of a single field mode. 

The most relevant difference between the toy model and the real case lies in the interaction term. We chose a smooth coupling function with a simple expression. This allowed for a simple perturbative treatment where we could obtain the near identity transformation analytically in detail. The real interaction term would be more complicated, since it involves the effective source described in previous sections. However, it can always be written as a functional of the action-angle variables that describe the particle's trajectory and the field configuration variables. In our peturbative treatment of the toy model we only considered the coupling between a single normal mode of the cavity and the oscillator. In the general case, the particle could couple to several (if not all) of the modes in the cavity at the same time. However as we have explored numerically the strength of the coupling is supressed by the physical distance between the orbital frequency $\Omega$ and the normal mode frequencies $\omega_{\ell \, n}$, and therefore only a few modes will contribute significantly. We have tested how this assumption modifies the results by checking numerically the perturbative results in comparison with a coupling function that introduced a coupling with different modes, showing that despite some quantitative differences, the qualitative behaviour is the same. 

In that sense, we argue that the set-up of a point-particle with a scalar-charge moving along geodesics of a cavity within the Schwarzschild metric can be effectively captured by a Hamiltonian which is essentially
\begin{equation}
    h = \sum_{\ell,m} h^{\ell m}_{\rm Oscillator} + h^{\ell m}_{\rm 1d-Cavity} + \frac{q}{m_0} \sum_n h^{\ell m}_{\rm Int, n},
\end{equation}
where the charge to mass ratio $q/m_0$ plays the role of the perturbative scale $\epsilon$, and $n$ describes the modes where the coupling induced by the effective source is relevant. This hamiltonian is subject to the same procedure in order to obtain a near-identity transformation that casts it into a trivial hamiltonian in a new set of dynamical variables. The procedure is particularly lengthy, especially due to the increased number of variables, but it is fundamentally the same. In particular one finds that there is a set of canonical variables $(Q_i,P_i)$ such that $Q_i = \omega_i(P_i) t +\dots$ for some momentum-dependent frequencies $\omega_i$ and $P_i = \rm const$. As a consequence, we can apply the same argument as we did for the toy model and conclude that the asymptotic value of the frequency of the charge, averaged in time, will be
\begin{equation}
    \Omega = \Omega_0 + \frac{q}{m_0}\delta^{(1)}\Omega + \left(\frac{q}{m_0}\right)^2 \delta^{(2)}\Omega,
\end{equation} 
where the first order term describes the exchange of energy between the initial configuration of the field inside the cavity $\Phi_0$ and the second order term also takes into account self-interactions. Since these are the asymptotic average states one should expect that it coincides with the frequency domain solution described in Section~\ref{sec:FD}. In particular, one can argue that by choosing a (weakly-populated) initial configuration $\Phi_0 = (q/m_0) \Psi_0$, it is always possible to choose this initial configuration such that it cancels out the $\delta^{(2)}\Omega$ term, so that the drift in frequency can be supressed. We conjecture that this argument could extend in higher order perturbation theory so that the drift can be arbitrarily supressed to any order by properly fine-tuning the initial conditions.

By studying the toy model we have also realized that there are chaotic configurations, which are related to (i) proximity resonant orbits and (ii) large couplings. Physically we are not considering the case where the coupling $q/m_0$ could be large. However, if we take into account that every angular mode acts like its own generalized version of the toy model that we discussed, it becomes increasingly harder to avoid initializing the system close to any resonant orbit. We have already described how requiring that the system has no resonances puts a constraint on the possible size of the cavity: in those situations, the system would never become chaotic. Since the perturbative scheme should produce similar results for the full case, we can estimate that the asymptotic drift in frequency is 
\begin{equation}
    D_\infty = \frac{q}{m_0} c_1 \Phi_0 + \left(\frac{q}{m_0}\right)^2 \frac{c_2}{(\Omega-\omega_n)^\alpha} + \dots,
\end{equation}
where $c_1$ and $c_2$ are some functions of the parameters of the problem, presumably of order $1$, $\omega_n$ is the closest resonant frequency and $\alpha\geq 1$ is some exponent. Typically, we can expect that the initial field configuration is supressed by the charge to mass ratio, $\Phi_0 = q/m_0 \Psi_0$. Then, chaos ensues once $D_\infty = \Omega-\omega_n$, i.e., when 
\begin{equation}
    \frac{q}{m_0} \sim (\Omega-\omega_n)^{\alpha/2} \, \sqrt[]{c_2 + c_1 \Psi_0}.
\end{equation}
Notice that if the initial conditions are fine-tuned to supress the drift $\Psi_0 = -c_2/c_1$, then the region in parameter space where chaos happens becomes smaller. However, for large cavities one would naturally expect that the system evolves through resonant orbits, which trigger the chaotic behaviour regardless of how small is the perturbative parameter. This could lead in general to escapes (the particle becomes unbound), or, most likely, to mergers (the particle's frequency becomes larger than the ISCO frequency).

\begin{figure}
    \centering
    \includegraphics[width = \columnwidth]{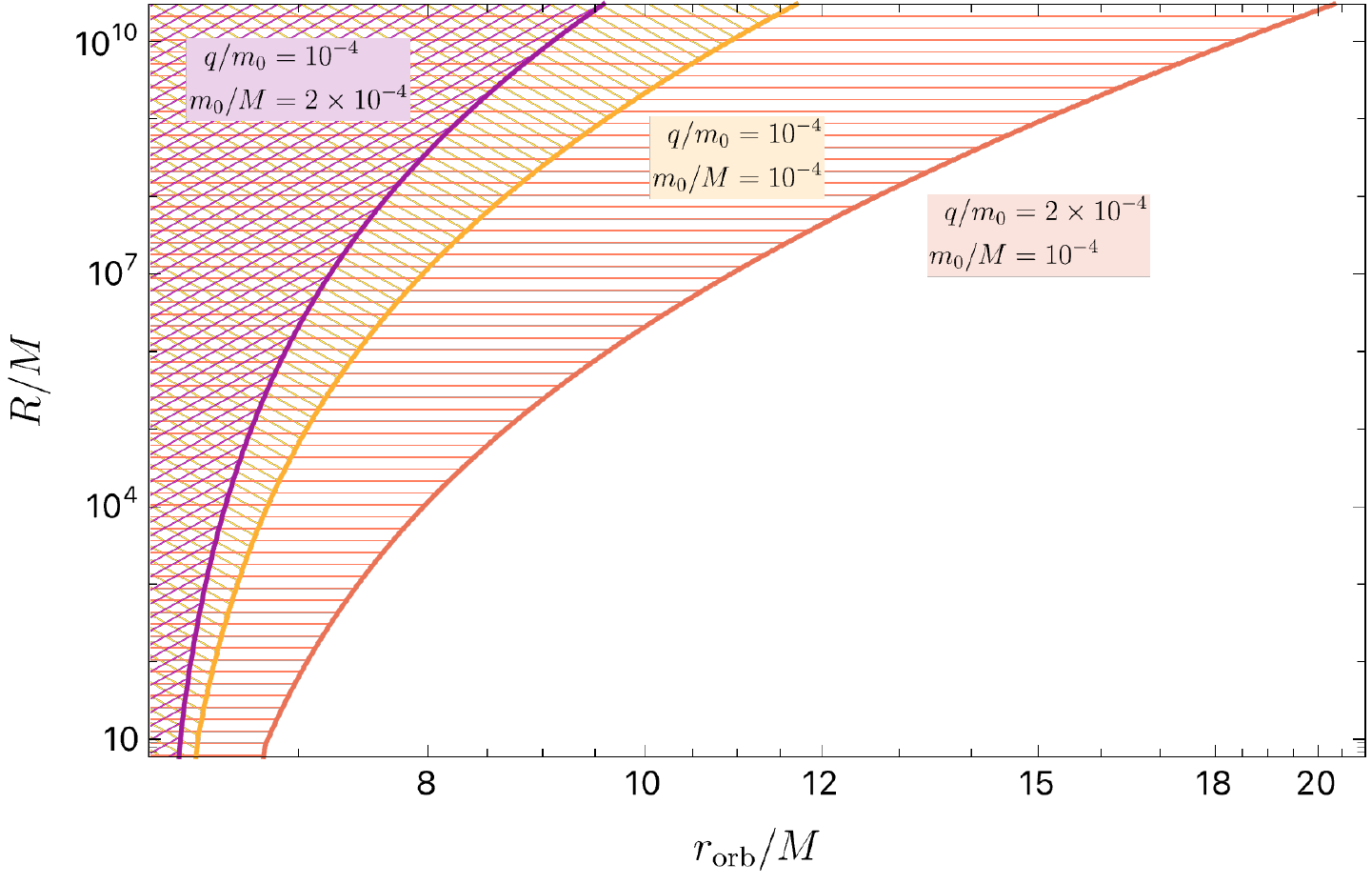}
    \caption{The shaded regions represent the configurations in terms of the distance between the orbital radius $r_{\rm orb}$ and the ISCO and the size of the cavity $R$ such that the orbits would be unstable due to exchanging enough energy with the cavity to merge before reaching equilibrium. We observe that, as intuitively expected, increasing the charge to mass ratio $q/m_0$ (c.f. yellow and orange regions) or decreasing the mass ratio $m_0/M$ (c.f. purple and yellow regions) results in a larger unstable region. }
    \label{fig:WhenUnstable}
\end{figure}

In the most general cases, a merger could happen preventing the system from reaching equilibrium. A rough way to estimate the likelihood of this phenomena is to compute the total energy contained in the asymptotic field configurations obtained from the frequency domain calculations. The energy can be expanded in angular modes as 
\begin{equation}\label{eq:energy-defn}
    \begin{aligned}
        E &= \sum_{\ell,m}\frac{1}{2}E_{\ell m}, \quad E_{\ell m} = \int_{r_1}^{r_2}dr \mathcal{E}_{\ell m}, \\
        \mathcal{E}_{\ell m} &= \left[(m\Omega)^2 + \frac{N\ell(\ell+1)}{r^2}\right](r \, \phi^{\ell m})^2 + N^2(r \, \partial_r\phi^{\ell m})^2.
    \end{aligned}
\end{equation}
The details of the calculation of the energy are discussed in Appendix~\ref{App:Energy}. In Tables~\ref{Table1}--~\ref{Table2} we compare the estimation of this energy with the energy that it would take to displace the original orbit towards the ISCO orbit $\Delta E$, defined as 
\begin{equation}\label{eq:Eorb-EISCO}
    \Delta E = |E(r_{\rm orb})-E(r_{\rm ISCO})|.
\end{equation}
We observe that, overall, the energy content in the field compared to the difference between the orbital energy and the ISCO energy is large. It is important to remark that this energy is supressed by the charge to mass ratio squared, which we assume to be very small. Moreover we can observe that this energy content increases with the size of the cavity: this is to be expected, as in the limit in which the cavity size is infinite we recover the asymptotically flat situation in which a merger is inevitable. 
\begin{table}[]
    \begin{tabular}{cc}
    \hline \hline
    $R/M$ & $(m_0/q)^2 E/\Delta E$ \\ \hline
    $10$  & $1654.72$              \\
    $12$  & $1754.67$              \\
    $14$  & $1817.61$              \\
    $16$  & $1861.96$              \\
    $18$  & $1895.46$              \\ 
    $20$  & $1922.08$              \\ \hline\hline
    \end{tabular}
    \caption{Values of the energy ratio $(m_0/q)^2 E/\Delta E$ between the late time energy content of the scalar field and the relative energy between the particle's orbit and the ISCO, for different cavity sizes $R/M$ at a fixed orbital radius $r_{\rm orb} = 7M$, where $\Delta E$ is given by Eq.~\eqref{eq:Eorb-EISCO} and measured per unit mass of the particle $m_0/M$. \label{Table1}}
\end{table}
\begin{table}[]
    \begin{tabular}{cc}
    \hline\hline
    $(r_{\rm orb}-r_{\rm ISCO})/M$ & $(m_0/q)^2 E/\Delta E$ \\ \hline
    $7$                            & $1754.67$              \\
    $8$                            & $672.72$               \\
    $9$                            & $420.10$               \\
    $10$                           & $309.23$               \\
    $11$                           & $241.29$               \\ 
    $12$                           & $183.51$               \\ \hline\hline
    \end{tabular}
    \caption{Values of the energy ratio $(m_0/q)^2 E/\Delta E$ between the late time energy content of the scalar field and the relative energy between the particle's orbit and the ISCO, as a function of the distance between the orbital radius $r_{\rm orb}$ and the ISCO radius, for a cavity with size $R = 12M$. As in the previous table, $\Delta E$ is measured per unit mass of the particle $m_0/M$. \label{Table2}}
    \end{table}
We have estimated the scaling of this quantity as a function of the cavity size and the distance between the orbit and the ISCO radius, as
\begin{equation}
    \frac{E}{\Delta E} \sim \frac{(R/M)^\alpha}{((r_{\rm orb}-r_{\rm ISCO})/M)^\beta}.
\end{equation}
The numerical values of $\alpha$ and $\beta$ obtained are:
\begin{equation}
    \begin{aligned}
        \alpha &= (0.20\pm 0.01), \\
        \beta &= (1.54 \pm 0.07).
    \end{aligned}
\end{equation}
Note that $\alpha$ and $\beta$ are consistent with $1/5$ and $3/2$, respectively. Now, taking into account that (i) the energy of the scalar field scales as the charge to mass ratio squared $(q/m_0)^2$, see Eq.~\eqref{eq:energy-defn} and (ii) that we have measured $\Delta E$ per unit mass of the point particle, we can estimate that the charge to mass ratio necessary to ensure that there is no merger before the system relaxes to equilibrium is given by 
\begin{equation}
    \frac{q}{m_0}\lesssim \sqrt{\frac{M}{m_0}} \left(\frac{R}{M}\right)^{\alpha/2} \left(\frac{M}{r_{\rm orb}-r_{\rm ISCO}}\right)^{\beta/2}.
\end{equation}
As we consider circular orbits closer to the ISCO, a smaller perturbative parameter $q/m_0$ will be sufficient to perturb the system. On the other hand, larger cavities can store more energy, resulting in configurations that could become unstable even for circular orbits far away from the ISCO. Finally it is interesting to note that for small mass ratios $m_0/M$, the charge is more easily displaced due to its own back-reaction, therefore it is sensible to expect that smaller charge to mass ratios $q/m_0$ would result in larger drifts and, potentially, unstable configurations. We can estimate the region in parameter space where a given orbit would become unstable. This is shown in Fig.~\ref{fig:WhenUnstable}, where we observe precisely this scaling. 
It is important to remark that this conclusion assumes that the initial field configuration is trivial. If the system is initialized with a scalar field with an energy similar to the value of the energy of the late time asymptotic configuration of the field, the drift induced due to the energy exchange between the particle and the field would be smaller. This is to say that by choosing appropiate initial conditions, the regions with unstable orbits can be made smaller. 

\section{Conclusions}
The dynamics of confined systems has shown to lead to interesting results in the past. The collapse of scalars fields in anti-de Sitter spacetimes, for example, may lead to collapse to black holes for a large class~\cite{Bizon:2011gg}, but not for generic initial data~\cite{Buchel:2013uba}. Here we focused on the two-body problem and shown that yet more surprises may be hidden within confined systems. We show that certain systems on circular orbits may be eternal, in truly confined systems, given appropriate initial conditions (see also~\cite{Dias:2012tq}). On the other hand, the presence of ``cavity'' modes leads to chaos in regimes of strong coupling or when the system is initialized close enough to a resonance. There results are mathematically interesting and relevant to gravitational systems such as anti-de Sitter spacetimes and possibly for binaries evolving within dark matter haloes, if it consists on massive fundamental fields.

We have assumed that the confined system is conservative, but in the context of massive degrees of freedom, this assumption is likely to fail, as one cannot prevent the radiation of gravitational waves. Thus, eternal binaries are clearly impossible once dissipation is allowed, but the transition to chaotic motion may still be present in full generality.

\acknowledgments
V.C.\ is a Villum Investigator and a DNRF Chair. We acknowledge financial support by the VILLUM Foundation (grant no. VIL37766) and the DNRF Chair program (grant no. DNRF162) by the Danish National Research Foundation. V.C. acknowledges financial support provided under the European Union’s H2020 ERC Advanced Grant “Black holes: gravitational engines of discovery” grant agreement
no. Gravitas–101052587. Views and opinions expressed are however those of the author only and do not necessarily reflect those of the European Union or the European Research Council. Neither the European Union nor the granting authority can be held responsible for them.
C.F.B.M. would like to thank
Fundação Amazônia de Amparo a Estudos e Pesquisas (FAPESPA), Conselho Nacional de Desenvolvimento Científico e Tecnológico (CNPq) and Coordenação de Aperfeiçoamento de Pessoal de Nível Superior
(CAPES) – Finance Code 001, from Brazil, for partial financial support.
This project has received funding from the European Union's Horizon 2020 research and innovation programme under the Marie Sklodowska-Curie grant agreement No 101007855.
We acknowledge financial support provided by FCT/Portugal through grants 
2022.01324.PTDC, PTDC/FIS-AST/7002/2020, UIDB/00099/2020 and UIDB/04459/2020.

\appendix
\section{Results in flat space}\label{App:Flat}
The fundamental physics describing the self-force of a charge in a circular orbit in Schwarzschild should not fundamentally depend on the structure of the metric. In fact we know that in the limit in which both the orbital radius and the location of the mirrors are far enough from the Schwarzschild radius the curvature of the spacetime will be small enough. In that regime, we expect results obtained by considering a Minkowski background to apply. These have the nice property of allowing for a mostly analytical analysis, as opposed to the situation for Schwarzschild spacetime. Here we derive the frequency domain self-force calculation replacing the Schwarzschild background for a Minkowski one, but keeping the problem otherwise unchanged. 

\subsection{Non-resonant regime}

\begin{figure}
    \centering
    \includegraphics[width = \columnwidth]{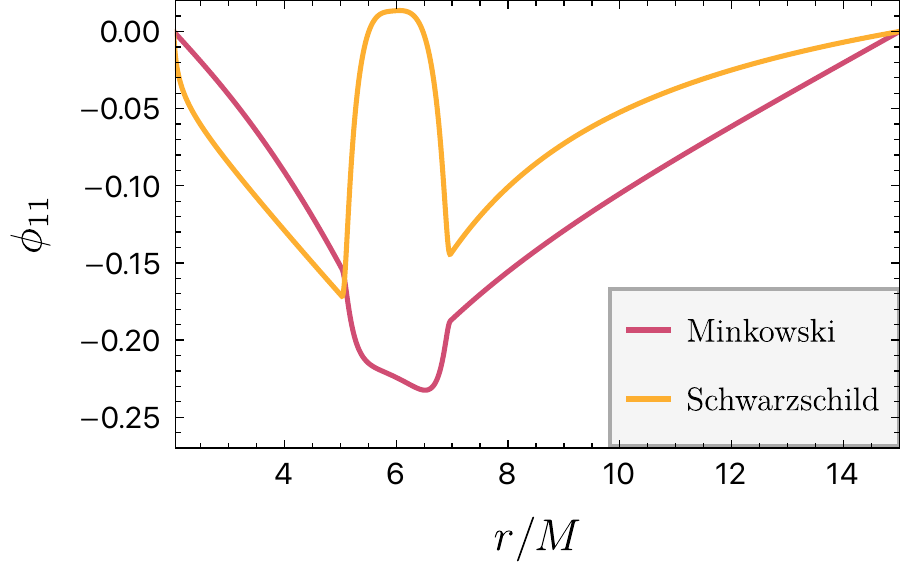}
    \caption{Regular field at the $(1,1)$ mode for a Schwarzschild cavity (yellow) and a Minkowski cavity (red), with mirrors placed ar $r_1 = 2.02M$ and $r_2 = 15M$ and the particle orbiting at $r_{\rm orb}=6M$. In the Minkowski cavity, even though $M=0$ we choose the same configuration of mirrors and orbital radius (and frequency) as in Schwarzschild. We observe that the asymptotic behaviour of the fields outside the region of the window function is most similar. The field close to the particle behaves in a very different way, due to the tail terms included in the puncture field for the Schwarzschild particle. }
    \label{fig:SchwarzschildVsMinkowski}
\end{figure}

We start by considering the non-resonant regime in detail. The Minkowski spacetime can be described by the same fundamental equation~\eqref{eq:KG_Schwarzschild}, but writing $f(r)=1$ and $V(r) = \ell (\ell+1)/r^2$. The homogeneous solutions are given by:
\begin{equation}
    \begin{aligned}
        \phi_{+(-)}(r) &= \sqrt{r}\left[J_{\ell+1/2}(\omega r)Y_{\ell+1/2}(\omega r_{2(1)})-\right.\\
        &\left.+J_{\ell+1/2}(\omega r_{2(1)})Y_{\ell+1/2}(\omega r)\right],
    \end{aligned}
\end{equation}
where $J$ and $Y$ are Bessel functions of the first and second kind. The puncture field is just given by the $M \to 0$ limit of Eq.~\eqref{eq:Puncture_Field}, 
\begin{equation}
    \phi_P^{\ell m} = Y_{\ell m}(\pi/2, 0)\frac{4\pi qr}{2r_{\rm orb}^2} \lvert r-r_{\rm orb}\rvert. 
\end{equation}
From this puncture field, obtaining the effective source is straightforward, and we can then solve for the field in exactly the same way as its done in the main text. In Fig.~\ref{fig:SchwarzschildVsMinkowski} we show, for comparison, the $(11)$ mode obtained for Schwarzschild and Minkowski, keeping the rest of the parameters identical. We observe that in the region where the only contribution is due to the homogeneous solution the two solutions are most similar. This is true despite considering a case in which the particle is exploring the strong field regime of the geometry. We observe clearly, though, that the field close to the particle behaves in a very different manner. This is due to the tail terms in~\eqref{eq:Puncture_Field} which are dominating in this regime.

%

\subsection{Resonant regime}
We can characterize explicitly the location of the resonances for the Minkowski cavity. For simplicity, set $r_1 = 0$ and $r_2 = R$. Then the inner solution is given by 
\begin{equation}
    J_-(r) = \sqrt[]{r}J_{\ell+1/2}(\omega r),
\end{equation}
and the Wronskian is just 
\begin{equation}
    \text{Wr}(\phi_+, \phi_-) = \frac{2}{\pi}J_{\ell+1/2}(\omega R).
\end{equation}
The normal modes are located at the zeros of the Wronskian. Therefore the frequencies of the normal modes can be characterized by 
\begin{equation}
    \omega_{\ell \, n} = \frac{j_{\ell+1/2 \, n}}{R}.
\end{equation}
We can estimate from this expression the minimum cavity size needed for a given frequency $\Omega$ to be resonant. The fundamental Bessel zero $j_{\ell \, 1} \sim \ell + \mathscr{O}(\ell)^{1/3}$~\cite[]{AbramowitzStegun}. Then, writing $\ell \Omega = \omega_{\ell \, 1}$ yields, for very large $\ell$:
\begin{equation}
    R_c = \frac{1}{\Omega}.
\end{equation}
Notice how this result coincides with the approximate scaling obtained for Schwarzschild.

\section{The vibrating string} \label{App:String}
The vibrating string is a simpler toy model that still captures most of the characteristics of the system that we are studying. In this section we will revisit this problem and solve it carefully using both the frequency domain and the Laplace transform approach. We observe explicitly that for closed systems the frequency domain calculation needs to be complemented with the excitation coefficients of the normal modes of the cavity at a given time. While in open systems the (quasi)-normal modes decay over time and therefore, after waiting a long enough period of time, the system achieves equilibrium, this is not the case for closed systems. A similar situation was observed recently for open systems with long-lived modes in~\cite{Cardoso:2022fbq}. 

In the following, we will be discussing a string with length $L$, which is forced with a frequency $\Omega$ at a particular point $x = x_0$. Its equation of motion is given by 
\begin{equation}
    -y_{,tt} + y_{,xx} = \delta(x-x_0)\cos(\Omega t), \quad y(0)=y(L) = 0.
\end{equation}
%
\subsection{Frequency domain approach}
Taking the Fourier transform of the above equation yields the inhomogeneous equation 
\begin{equation}
\psi_{,xx} + \omega^2 \psi = \delta(x-x_0)\delta(\omega-\Omega),
\end{equation}
where $\psi(x)$ is the Fourier transform of the string amplitude $y$. It is not hard to check that the solution to this equation is given by 
\begin{equation}
    \begin{aligned}
        \psi &= \frac{\delta(\omega-\Omega)}{\omega\sin(\omega L)}\left[\sin(\omega x)\sin(\omega(L-x_0))\theta(x_0-x) + \right.\\
        &\left.+\sin(\omega x_0)\sin(\omega(L-x))\theta(x-x_0)\right],
    \end{aligned}
\end{equation}
where $\theta(x)$ is the Heaviside step function. When transforming back to time domain we arrive at the solution 
\begin{equation}
    \begin{aligned}
        y_{\rm FD} &= \frac{\cos(\Omega t)}{\Omega\sin(\Omega L)}\left[\sin(\Omega x)\sin(\Omega(L-x_0))H(x_0-x) + \right.\\
        &\left.+\sin(\Omega x_0)\sin(\Omega(L-x))H(x-x_0)\right].
    \end{aligned}
\end{equation}
The string then oscillates with a single frequency $\Omega$, which coincides with the driving frequency (which we assume to not be resonant for simplicity, i.e., $\Omega\neq n \pi /L$ for $n\in\mathbb{Z}$). Adding any linear combination of homogeneous solutions (of normal modes) would still be a solution to the problem. However, the frequency domain calculation is not informative about the excitation coefficients of these normal modes. In order to obtain these we need to consider a slightly different approach.

\subsection{Laplace transform}
Studying the Laplaced transform system will allow us to obtain the excitation coefficients in terms of the characteristics of the initial conditions. In order to allow for more generality, we consider an arbitrary source profile $S(x)$. 
\begin{equation}
    \varphi_{,xx} -s^2\varphi = S(x)\frac{s}{s^2+\Omega^2}-su(x)-v(x),
\end{equation}
where $y(t=0,x)=u(x)$ and $y_{,t}(t=0,x)=v(x)$ are the initial displacement and velocity, respectively. We can directly write the solution to this equation with Dirichlet boundary conditions as 
\begin{equation}\label{eq:Laplace-Sol}
    \varphi = \int_0^L dy F(s,y) G(s,x,y) + \int_0^x dy F(s,y)H(s,x,y),
\end{equation}
where
\begin{equation}
    \begin{aligned}
        F(s,x)&= S(x)\frac{s}{s^2+\Omega^2}-su(x)-v(x), \\
        G(s,x,y) &= \frac{1}{2s}\frac{e^{sx}-e^{-sx}}{e^{-sL}-e^{sL}}\left(e^{s(L-y)}-e^{-s(L-y)}\right), \\
        H(s, x, y)&= \frac{1}{2s}\left(e^{s(x-y)}-e^{-s(x-y)}\right).
    \end{aligned}
\end{equation}
Since both $G$ and $H$ are solutions of the homogeneous equation with respect to the variable $x$, it is easy to check that this is indeed a solution to the Laplace transformed equation. Notice that $H(s,x,y)$ is holomorphic in $s$~\footnote{The apparent singularity at $s=0$ is regularized by the term between brackets, as can be seen by simply applying L'Hopital's rule.}, whereas $G(s,x,y)$ is meromorphic: it has simple poles at the normal modes of the cavity $s_k =i\omega_k = i k \pi/L$ for $k$ a non-zero integer number. Transforming back to time domain now is more complicated than in the frequency domain case. However, it is still possible to obtain simple analytical solutions in this case. In the more general case, where the structure of the poles is less clear, numerical approaches to the inverse Laplace transform are also possible. 

\begin{figure}
    \centering
    \includegraphics[width = \columnwidth]{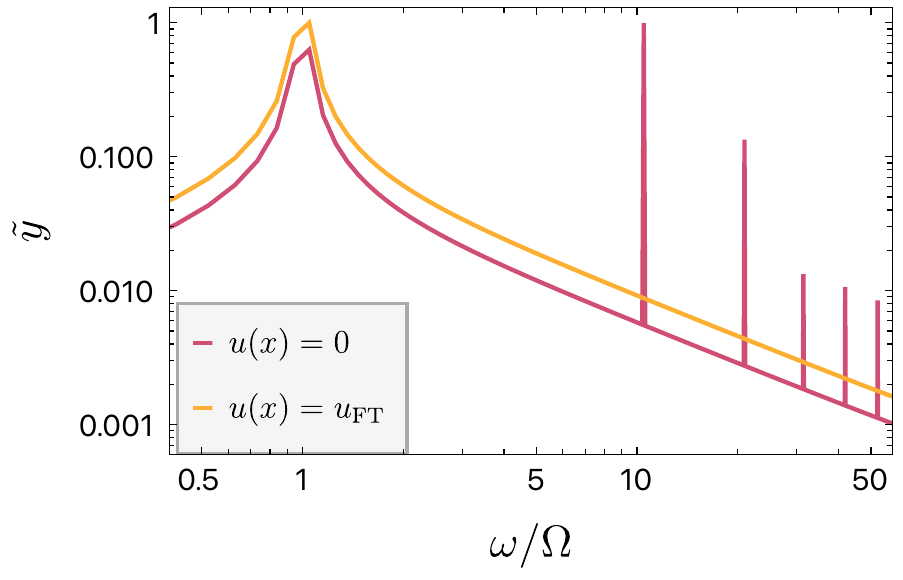}
    \caption{Fourier transform $\Tilde{y}$ of the displacement of the string at $x=0.3L$ for a sinusoidal source with $\Omega=0.3$ and $L=1$. We consider both trivial ($u(x)=0$) and fine-tuned ($u(x)$ as given in Eq.~\eqref{eq:fine-tuning-string}) initial conditions. }
    \label{fig:SFString}
\end{figure}

In order to analytically invert the Laplace transform we will make use of the convolution theorem. Then, we can write the solution as 
\begin{equation}
    y(t,x) = \int_0^L dy \int_0^t d\tau f(\tau, y)g(t-\tau, x, y),
\end{equation}
where $f(t, x) = \mathcal{L}^{-1}[F(s,x)]$ and $g(t, x, y) = \mathcal{L}^{-1}[e^{st}G(s,x)]$. There is no contribution from the second term in~\eqref{eq:Laplace-Sol} since the inverse transform of $H$ vanishes. The first term is straightforward:
\begin{equation}
    f(t,x) = S(x)\cos(\Omega t) - \delta(t)v(x) - \delta_{,t}(t)u(x).
\end{equation}
The second term involves computing the Bromwich integral. However since all the poles are simple, we can write the integral as the sum over the residues using Cauchy's theorem:
\begin{equation}
    g(t,x,y) = -\sum_{k\in\mathbb{Z}-\{0\}} \frac{1}{L\omega_k}\sin(\omega_k x)\sin(\omega_k y)\sin(\omega_k t).
\end{equation}
Finally computing the convolution integral and defining the normal mode coefficients of any function $A(x)$ as 
\begin{equation}
    A_k = \frac{1}{L}\int_0^L dy A(y)\sin(\omega_k y)
\end{equation}
yields the result 
\begin{equation}
    \begin{aligned}
        y(t, x) &= \sum _{k\in\mathbb{Z}-\{0\}} \left[u_k\cos(\omega_k t) + \frac{v_k}{\omega_k} \sin(\omega_k t)\right.\\
        &\left.+ S_k\frac{\cos(\Omega t)-\cos(\omega_k t)}{\Omega^2-\omega_k^2}\right]\sin(\omega_k x).
    \end{aligned}
\end{equation}
It is clear from this expression that there are initial configurations $\{u_n\}$ such that the average displacement during a driving period (i.e. all of the normal mode contributions, except for the one associated to the driving frequency), vanish. It is enough to choose 
\begin{equation}\label{eq:fine-tuning-string}
    u_k = \frac{S_k}{\Omega^2-\omega_k^2}, \quad v_k = 0,
\end{equation}
to obtain this behaviour, see Fig.~\ref{fig:SFString}. The string displacement has a single frequency peak at the orbital frequency $\Omega$ for the fine tuned initial conditions, whereas if the initial condition is just zero there are multiple peaks at the normal modes that are excited by the source. 
Finally, the solution obtained through the Laplace transform has a regular limit at the resonances. Just taking the $\Omega\to \omega_k$ limit yields 
\begin{equation}
    \begin{aligned}
        y(t, x) &= \sum _{k\in\mathbb{Z}-\{0, k_r\}} \left[u_k\cos(\omega_k t) + \frac{v_k}{\omega_k} \sin(\omega_k t)\right.\\
        &\left.+ S_k\frac{\cos(\Omega t)-\cos(\omega_k t)}{\Omega^2-\omega_k^2}\right]\sin(\omega_k x) - \\
        &- \frac{S_{k_r}}{\omega_{k_r}} t \sin(\omega_{k_r}t) \sin(\omega_{k_r} x).
    \end{aligned}
\end{equation}
We observe that the resonant mode $k_r$ grows linearly in time. Since the energy in the string is quadratic in the amplitude, this means that the energy would grow quadratically too, consistently with~\cite{Annulli:2020lyc}.

\section{Near-Identity Transformation}\label{App:Transformation}
In this Appendix we construct the near-identity transformation that will help us solve perturbatively the toy model described in the main text. We follow Chapter 5 of~\cite[]{kevorkian2012multiple}. In particular, we consider a hamiltonian
\begin{equation}
    h = \frac{p_1^2}{2} + \omega p_2  - \epsilon \, \, \sqrt[]{\frac{2 s p_2}{\omega}} \cos q_1 \cos q_2.
\end{equation}
We write this now in terms of the canonical variables ($(q_1,q_2,p_1,p_2) = (q, \varphi, \Omega, J)$ in the previous notation), and $\omega$ is the (constant) frequency of the normal mode of the cavity that couples to the oscillator. We will decompose each variable into its average part $\bar{f}$ and its oscillatory part $\check{f}$, where 
\begin{equation}
    \bar{f} = \frac{1}{(2\pi)^2}\int dq_1dq_2 f,
\end{equation}
and $\check{f} = f - \bar{f}$. The goal is to find a generating function, which we expand as 
\begin{equation}
    F(q_i, P_i, \tilde{t}) = q_i P_i + \epsilon(\bar{F}_1 + \check{F}_1) + \epsilon^2 F_2 + \mathscr{O}(\epsilon)^3.
\end{equation}
such that the transformed hamiltonian is trivial. This transformed hamiltonian, in general, is given by 
\begin{equation}
    H = h - \epsilon \frac{\partial F}{\partial \tilde{t}} = H_0 + \epsilon H_1 + \epsilon^2 H_2 + \mathscr{O}(\epsilon)^3,
\end{equation}
where $\tilde{t} = \epsilon t$ is a slow time. This variable is introduced to make sure that the solution is valid up to the given order in $\epsilon$ at all times. Notice that a naive perturbative analysis would yield a solution which is only valid up to $0<t<1/\epsilon$ (to first order). The first order contribution to the new Hamiltonian is given by 
\begin{equation}
    H_1 = \check{h}_1 + \omega_j \frac{\partial F_1}{\partial q_j},
\end{equation}
where $\omega_j = (p_1, \omega)$ is just the derivative of the un-perturbed angle variables. Notice that the secular part of $F_1$ does not enter in this term since the system is hamiltonian. Now requiring that this vanishes yields the following condition for the oscillatory part of $F_1$:
\begin{equation}
    \check{F}_1 = -\left[\int d\tau \check{h}_1(P_i, \omega_i s)\right]\rVert_{\omega_i \tau = q_i},
\end{equation}
which for our case is simply written as 
 \begin{equation}
    \check{F}_1 = \sqrt{\frac{2sP_2}{\omega}}\frac{P_1\sin q_1 \cos q_2 - \omega \cos q_1 \sin q_2}{P_1^2 - \omega^2}.
 \end{equation}
The second order contribution to the hamiltonian in our case is now
\begin{equation}
    \begin{aligned}
        H_2 &= \frac{\partial F_1}{\partial \tilde{t}} + \omega_j \frac{\partial F_2}{\partial q_j} - Z, \\
        Z &= \frac{1}{2}\frac{\partial \omega_j}{\partial p_k}\frac{\partial \check{F}_1}{\partial q_j}\frac{\partial \check{F}_1}{\partial q_k} - \omega_k \frac{\partial^2 \check{F}_1}{\partial q_k \partial P_j}\frac{\partial \check{F}_1}{\partial q_j}.
    \end{aligned}
\end{equation}
We also want to cancel this term. We can use the secular part of $F_1$ to cancel the secular contribution to this hamiltonian, and the oscillatory part of $F_2 = \check{F}_2$ to cancle the oscillatory part. Summing up, these are the two conditions that we need to satisfy:
\begin{equation}
    \begin{aligned}
        \omega_j \frac{\partial F_2}{\partial q_j} &= -\frac{\partial \check{F}_1}{\partial \tilde{t}} + \check{Z}, \\
        \frac{\partial \bar{F}_1}{\partial \tilde{t}} &= \bar{Z}.
    \end{aligned}
\end{equation}
In order to solve these equations, we first write down the average and oscillatory parts of $Z$ erxplicitly:
\begin{equation}
    \begin{aligned}
        \bar{Z} &= -\frac{s \left[(P_2 - \omega)\omega^2 + P_1^2(P_2 + \omega)\right]}{4(P_1 - \omega)^2 \omega(P_1 + \omega)^2}, \\
        \check{Z} &= \frac{s}{4\omega(P_1^2 - \omega^2)^2}\left[c_1 \cos^2q_1\cos^2q_2 + \right.\\
        &\left.+c_2 \sin^2q_1\sin^2q_2 + c_3 \sin(2q_1)\sin(2q_2)\right], \\
        c_1 &= \omega^2 (P_2 - \omega) + P_1^2 (P_2 + \omega) - 4\left[P_1^2 (P_2 + \omega)\right], \\
        c_2 &= -4P_2\omega^2, \\
        c_3 &= P_1\left[\omega(\omega - 2P_2) - P_1^2\right].
    \end{aligned}
\end{equation}
The second of the equations can be solved immediately, since $\bar{Z}$ does not depend on time, so 
\begin{equation}
    \bar{F}_1 = \bar{Z} \tilde{t}.
\end{equation}
and we can directly solve the first equation:
\begin{equation}\label{eq:F2-NIT}
    \begin{aligned}
        F_2 &= \frac{s}{8P_1\omega^2(P_1-\omega)^3 (P_1 + \omega)^3}\left[d_1\sin(2q_1) + \right.\\
        &\left.+d_2\sin(2q_2) + d_3 \sin(2q_1)\cos(2q_2) + d_4 \cos(2q_1)\sin(2q_2)\right], \\
        d_1 &= \omega (P_2 + \omega)(P_1^2 - \omega^2)^2, \\
        d_2 &= -P_1 (P_2 + \omega)(P_1^2 - \omega^2)^2, \\
        d_3 &= -P_1^2\omega \left[(3P_2-2\omega)\omega^2 + P_1^2(P_2+2\omega)\right], \\
        d_4 &= \omega\left[ P_1^4 + 3P_1^2P_2\omega + (P_2-\omega)\omega^3 \right].
    \end{aligned}
\end{equation}
This completely characterizes the near-identity transformation. Notice that a second-order accurate solution in the momenta does not require us to fix the secular part of $F_2$, which we set to zero for simplicity. 

\section{Numerics}\label{App:Numerics}
We solve the system in first order form~\eqref{eq:toy-model-first-order} numerically using the method of lines. We implement an overall fourth order discrete scheme using summation by parts operators as well as fourth order Kreiss-Oliger dissipation. The integrals are evaluated using Simpson's method: despite a lower accuracy than other adaptative integration schemes, this allows for a faster implementation than methods requiring extrapolation. 
The time-integration is implemented via a fourth-order Runge-Kutta scheme with an adaptative step-size. With this set-up we are able to evolve the system in a stable fashion. Since the relaxation to equilibrium of the system happens very quickly it is not necessary for our purposes to implement time integrators adapted to energy conservation (such as symplectic ones). 

\begin{figure}
    \centering
    \includegraphics[width = \columnwidth]{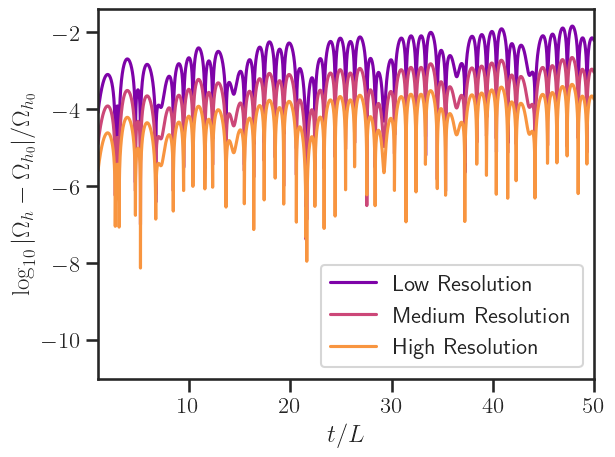}
    \caption{Relative difference between the evolution of the frequency using three different resolution scales $h = 10h_0, 4h_0, 2h_0$ respectively, with the evolution of the frequency using a discretization scale of $h_0 = 5 \times 10^{-3}L$. Increasing the resolution decreases the relative difference, consistently with fourth order convergence. }
    \label{fig:Convergence}
\end{figure}

We study the convergence by comparing the relative difference between a run with a very high resolution $h_0 = 5 \times 10^{-3}L$ in the spatial discretization with three worse resolutions. We show in Fig.~\ref{fig:Convergence} that increasing the resolution improves uniformly the agreement. This guarantees the convergence of our numerical scheme. 
%
\section{Energy in the cavity}\label{App:Energy}
\begin{figure}
    \centering
    \includegraphics[width=0.99\columnwidth]{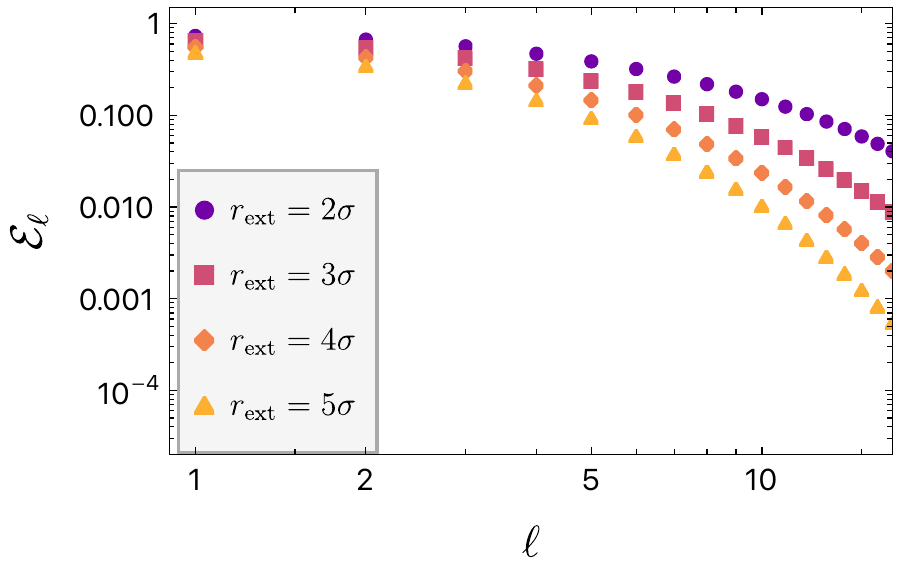}
    \caption{Energy of the modes $\mathcal{E}_{\ell m}$ as a function of the angular number $\ell$, for different values of the extraction radius $r_{\rm ext}$. Clearly, the convergence is faster as the radius of the excision is larger. We find that extracting at $r_{\rm ext} = 3\sigma$ provides a quick convergence and stable results.}
    \label{fig:Energy-Modes}
\end{figure}

In this Appendix we estimate the energy contained in the scalar field configuration obtained from the Frequency Domain calculation. This is an interesting way of estimate the final state of the system: whether in the process of approaching equilibrium the charge was merged into the central compact object before achieving equilibrium. The energy density of the field is given by 
\begin{equation}
    \begin{aligned}
        \rho =& T^{00} = \frac{N}{2}\left[\frac{1}{N} (\partial_t\Phi)^2 + N (\partial_r \Phi)^2\right.\\
        &\left.+ \frac{1}{r^2}(\partial_\theta \Phi)^2+\frac{1}{r^2\sin^2\theta}(\partial_\varphi \Phi)^2\right].
    \end{aligned}
\end{equation}
We will use the spin raising and lowering operators $\eth$ and $\bar{\eth}$~\cite{Goldberg:1966uu}, which act on spin-weighted spherical harmonics as 
\begin{equation}
    \begin{aligned}
        \eth Y^{(s)}_{\ell m} &= \sqrt[]{(\ell-s)(\ell+s+1)}Y^{(s+1)}_{\ell m}, \\
        \bar{\eth} Y^{(s)}_{\ell m} &= \sqrt[]{(\ell+s)(\ell-s+1)}Y^{(s-1)}_{\ell m},
    \end{aligned}
\end{equation}
where $(s)$ is the spin weight of the spherical harmonics. When acting on a quantity with spin weight $s = 0$ the spin raising and lowering operators are just given by 
\begin{equation}
    \eth S = \partial_\theta S + \frac{i}{\sin\theta}\partial_\varphi S,
\end{equation}
so we can write the above energy density as 
\begin{equation}
    \rho = \frac{N}{2}\left[\frac{1}{N}(\partial_t\Phi)^2 + N(\partial_r\Phi)^2 + \frac{1}{r^2}\eth\Phi\bar{\eth}\Phi\right].
\end{equation}
We integrate the energy density on the cavity to obtain the total energy
\begin{equation}
    E = \int_{r_1}^{r_2}r^2 dr \int_{S^2} \rho = E_t + E_r + E_{\rm Ang}.
\end{equation}
The temportal part taking into account that in the asymptotic stationary state each field oscillates with a frequency $m\Omega$ is given by 
\begin{equation}
    \begin{aligned}
        2E_t &=\sum_{\ell,m,L,M} \int_{r_1}^{r_2}r^2 dr\partial_t\phi^{\ell m}  \partial_t\phi^{LM} \int_{S^2} Y_{\ell m}Y_{LM} \\
        &= \sum_{\ell,m} (m\Omega)^2 \int_{r_1}^{r_2}dr (r\, \phi^{\ell m})^2,
    \end{aligned}
\end{equation}
where we have used the orthonormality of the spherical harmonics and the partiy of the field. Using similar arguments, the radial contribution to the energy is 
\begin{equation}
    E_r = \frac{1}{2} \sum_{\ell, m}\int dr (r \, N \, \partial_r \phi_{\ell m})^2.
\end{equation}
Finally the angular part is given by 
\begin{equation}
    \begin{aligned}
        2E_{\rm Ang} &= \sum_{\ell,m,L,M}\int_{r_1}^{r_2} r^2 dr \frac{N}{r^2}\phi^{\ell m}\phi^{LM} \int_{S^2}\eth Y_{\ell m}\bar{\eth}Y_{LM} \\
        &= \sum \ell(\ell+1)\int_{r_1}^{r_2}dr N \phi_{\ell m}^2.
    \end{aligned}
\end{equation}
Putting everything together, the total energy is 
\begin{equation}
    \begin{aligned}
        E &= \frac{1}{2}\sum_{\ell,m} \int_{r_1}^{r_2}dr\mathcal{E}_{\ell m}, \\
        \mathcal{E}_{\ell m} &= \left[(m\Omega)^2 + \frac{N\ell(\ell+1)}{r^2}\right](r \, \phi^{\ell m})^2 + N^2( r\, \partial_r\phi^{\ell m})^2.
    \end{aligned}
\end{equation}
We can compute this energy for the stationary configurations obtained from the frequency domain approach. We evaluate the radial integral numerically, first obtaining a sufficiently smooth interpolator for the field. In order to ensure mode convergence, we excise a region around the particle of radius $r_{\rm ext} \propto \sigma$, where $\sigma$ is the width of the Gaussian window function used to describe the puncture field. This way, we effectively remove the contribution to the energy contained in the cavity due to the local field. In Fig.~\ref{fig:Energy-Modes} we show that it is necessary to exclude a region of a radius of $r_{\rm ext} \geq 3\sigma$ in order to achieve convergence of the mode sum of the energy.

\bibliography{ref}


\end{document}